# A Comprehensive Review of One-Pixel Attack: Research Status, Taxonomy, Applications, Regulation Policy and Future Directions

**Mirza Niaz Morshed**
mirza.n.morshed@gmail.com
ORCID: 0009-0005-7918-9604
Bangladesh University of Business and Technology, Dhaka, Bangladesh

**Md. Masudul Islam**
masudulislam11@gmail.com
ORCID:  0000-0001-7643-5420
Bangladesh University of Business and Technology, Dhaka, Bangladesh

**Galib Muhammad Shahriar Himel**
galib.muhammad.shahriar@gmail.com
ORCID: 0000-0002-2257-6751
Universiti Sains Malaysia, Pulau Pinang, Malaysia

**Md. Aslam Uddin**
aslam@bubt.edu.bd
ORCID: 0000-0003-2994-1317
Bangladesh University of Business and Technology, Dhaka, Bangladesh

**Hui Liu**
hui.liu@uni-bremen.de
ORCID: 0000-0002-6850-9570
Universität Bremen, Germany

**Md. Shafiqul Islam**
msislam@uap-bd.edu
ORCID: 0000-0001-9469-2041
University of Asia Pacific, Dhaka, Bangladesh

***Corresponding author:** Md. Masudul Islam (E-mail: masudulislam11@gmail.com)

## Abstract

One-Pixel Attacks (OPAs) represent one of the most extreme demonstrations of adversarial fragility in deep learning, where modifying a single pixel can reliably induce high-confidence misclassification across domains such as medical diagnosis, autonomous driving, biometrics, and quantum communication. Despite their conceptual simplicity, OPAs remain underexamined in existing adversarial-attack surveys, which provide only fragmented or cursory coverage. This PRISMA-guided review synthesizes high-quality studies from 2017 to 2026 and delivers a unified, multi-axis taxonomy of OPA research spanning algorithmic foundations, black-box evolutionary optimization, emerging hybrid and program-synthesis attacks, defence mechanisms, interpretability tools, and domain-specific vulnerabilities. Our analysis reveals the dominance of Differential Evolution–based strategies, the rise of efficiency-optimized and saliency-guided methods, and persistent gaps in dataset diversity, transferability, and standardized evaluation. We summarized and assess defence paradigms including pixel restoration, anomaly detection, input-space transformations, and robust training highlighting their trade-offs in robustness, imperceptibility, and computational overhead. Building on these insights, we outline future research priorities involving selective pixel recovery, transformer-specific vulnerability analysis, saliency-driven optimization, and real-world domain-adaptive defences. We further propose a regulatory framework emphasizing robustness testing, incident disclosure, and AI security governance. This review establishes a comprehensive foundation for understanding, evaluating, and mitigating ultra-sparse adversarial threats in contemporary AI systems.

**Keywords:** Adversarial Attack, One Pixel Attack, AI Vulnerability, Computer Vision, AI Security.

## 1. Introduction

Artificial intelligence has achieved remarkable success in healthcare, transportation, biometrics, industrial automation, and scientific imaging, making the reliability of these systems increasingly vital. As AI models play a growing role in high-stakes decision-making, ensuring their robustness and trustworthiness has become essential. Deep neural networks (DNNs) are the main technology driving many of these AI systems, yet their impressive performance is offset by a notable vulnerability to adversarial perturbations, which poses a fundamental security concern. Even small, carefully crafted modifications can cause models to make incorrect predictions with high confidence, highlighting weaknesses in their learned decision boundaries [1][2]. Among the broad spectrum of adversarial techniques, the one-pixel attack (OPA) introduced by Su et al. [3] is particularly striking: by altering a single pixel, an attacker can reliably mislead a classifier without producing perceptible changes to the image. This extreme sparsity makes OPA one of the simplest yet most revealing tools for analysing neural network fragility.

The disruptive potential of OPA has been demonstrated across several safety-critical contexts. In medical imaging, even a single-pixel perturbation can shift diagnostic predictions, raising concerns about the reliability of automated decision-support systems [4]. In autonomous driving, OPAs have been shown to alter the recognition of traffic signs, creating risks for perception modules in vehicles [5]. Similarly, minimal perturbations have been used to deceive biometric authentication systems, exposing vulnerabilities in face-recognition pipelines [6]. Research has also highlighted implications for quantum imaging and remote sensing, where high-precision measurements can be distorted by ultra-sparse adversarial noise [7]. These findings collectively emphasize that even minimal adversarial manipulation can have disproportionate real-world consequences.

Despite the significance of OPA, it remains underexplored in existing literature. Most adversarial machine-learning surveys concentrate on gradient-based or ℓp-bounded perturbation methods, including FGSM [8], PGD [9], DeepFool [10], and the C&W attack [11]. In these reviews, OPA is typically mentioned only briefly, often as an early example of black-box or query-efficient attacks, without detailed analysis of its optimization strategies, algorithmic variants, defence mechanisms, or domain-specific implications. As a result, there is currently no systematic review that consolidates knowledge on OPA or examines its evolution, strengths, limitations, and real-world risks.

This gap in the literature highlights the need for a focused and structured synthesis. Understanding OPA is especially important because its minimal adversarial footprint provides unique insights into pixel-level saliency, boundary sensitivity, and model interpretability insights that are not always accessible through conventional adversarial methods. To address this need, our study presents the PRISMA-guided systematic review dedicated exclusively to one-pixel and ultra-sparse adversarial attacks, covering research published between 2017 and 2026. To ensure a systematic synthesis, this study adopts the PICO framework [12] to structure the review along the following core dimensions:

i. **P (Problem):** How do one-pixel adversarial attacks impact different computer vision models, including CNNs, domain adapted architectures, and applications in fields such as medical imaging, industrial IoT, and security?
ii. **I (Intervention):** What are the characteristics and effectiveness of one-pixel and related multi-pixel adversarial methodologies, as well as notable defence mechanisms and detection strategies?
iii. **C (Comparison):** How do these attacks and defences compare to clean (unaltered) inputs, alternative attack types, or traditional defence mechanisms?
iv. **O (Outcome):** What are the main outcomes in terms of attack success rate, model performance degradation, and the robustness and efficiency of various defence strategies across domains?

Based on these elements, our review is driven by the following objectives:

- To systematically examine the success rates, characteristics, and limitations of OPA across diverse computer vision tasks and domains;

- To critically evaluate and compare the effectiveness and trade-offs of available defence strategies;
- To identify domain-specific vulnerabilities and practical challenges in deploying robust models;
- To highlight current research gaps and propose evidence-based directions for further exploration and methodological development.

To fulfil these objectives, this paper makes the following core contributions:

- We integrate findings from all possible studies across 2017–2026 and consolidate them into a multi-dimensional framework covering attack settings, optimization strategies, model families, evaluation protocols, and domain contexts. This goes beyond cataloguing prior works by revealing methodological patterns, shared assumptions, and recurring weaknesses across the OPA landscape.
- By analysing OPA research across natural images, medical imaging, biometrics, autonomous systems, and emerging security domains, we identify how vulnerabilities and robustness vary by application type. This comparative perspective clarifies where OPAs remain largely theoretical versus where they present concrete real-world risks.
- Using structured meta-analysis from our review dataset, we quantify the field's progress and highlight structural gaps, including limited dataset diversity, overuse of toy benchmarks, and inconsistent evaluation metrics. This evidence-driven critique helps position OPA research within broader adversarial machine learning maturity levels.
- Building on the insights above, we propose an OPA-specific governance and risk-management model encompassing developers, deployers, auditors, and regulators. This framework outlines practical guidelines for responsible testing, robustness disclosure, model documentation, and domain-specific safety thresholds.

## 2. Background and Theoretical Foundation of OPA

OPAs expose a fundamental vulnerability in deep neural networks (DNNs) even a single strategically altered pixel can force misclassification. Unlike traditional adversarial attacks [116] bounded by $(L_2)$ or $(L_\infty)$ norms, OPAs fall under $(L_0)$-norm perturbations, where the goal is to minimize the number of altered features rather than their magnitude. This makes the optimization problem inherently discrete and combinatorial, often motivating the use of black-box search methods [13] such as Differential Evolution (DE) [14] (Price, 2013).

In high-dimensional input spaces $(R^n)$, decision boundaries of deep models tend to be highly non-linear and frequently lie close to natural data points. Consequently, perturbing a single pixel $\left(x_i \to x_i'\right)$ can shift the overall input $(x \to x')$ across the decision boundary which is $\left(f(x) \neq f(x')\right)$ even when $(|x - x'|_0 = 1)$. This effect is amplified when the perturbed pixel has high saliency, meaning it strongly contributes to the model's internal feature activations. In low-resolution images or shallow architectures, where each pixel carries greater semantic weight, this vulnerability becomes even more pronounced. **Figure 1** (a) depicts this geometric intuition, and **Figure 1** (b) shows the search space explored by single- and two-pixel attacks. A more abstract depiction of the OPA mechanism is presented in **Figure 2**, highlighting how a seemingly negligible perturbation can drastically change a model's prediction. The mathematical Formulation can be expressed as,

Let $\left(x \in [0,1]^{H\times W\times C}\right)$ denote a clean input image and $\left(f(\cdot)\right)$ a classifier producing the predicted label $(f(x) = y)$. A one-pixel perturbation $(\delta)$ has support restricted to a single pixel position (p) that is $(|\delta|_0 \leq 1)$. The adversarial optimization can be expressed as:

$$\left[\quad L\left(f\left(x+\delta_{p,v}\right),\ \ y_{target}\right) \quad or \quad L\left(f\left(x+\delta_{p,v}\right),\ y\right),\right]$$

subject to

$$\left[\delta_{p,v}(i) = 0; \forall i \neq p, \qquad v_{min} \leq v \leq v_{max}\right],$$

$\left(\delta_{p,v}\right)$ assigns pixel (p) a new color value (v), and $(L)$ is a loss such as cross-entropy. Solving this non-convex optimization typically relies on population-based search (DE, PSO, CMA-ES), greedy/critical-pixel strategies, or program-synthesis-based methods. These differ in query complexity, transferability, and perceptual imperceptibility. Best-practice evaluation for OPAs generally reports: (i) Attack Success Rate (ASR) [15] ii. query count/runtime [16] iii. perceptual similarity [17], PSNR [18], SSIM [19] iv. confidence shift in the model output.

Together, these metrics align with the optimization objectives and ensure reproducible comparisons across studies. OPAs therefore represent the most minimal and interpretable form of adversarial vulnerability. They provide unique insight into decision boundary fragility, pixel-level salience, and structural weaknesses in both CNNs and transformer-based architectures. Their implications for reliability in real-world AI systems including medical imaging, biometrics, and autonomous perception are illustrated in **Figure 3**.

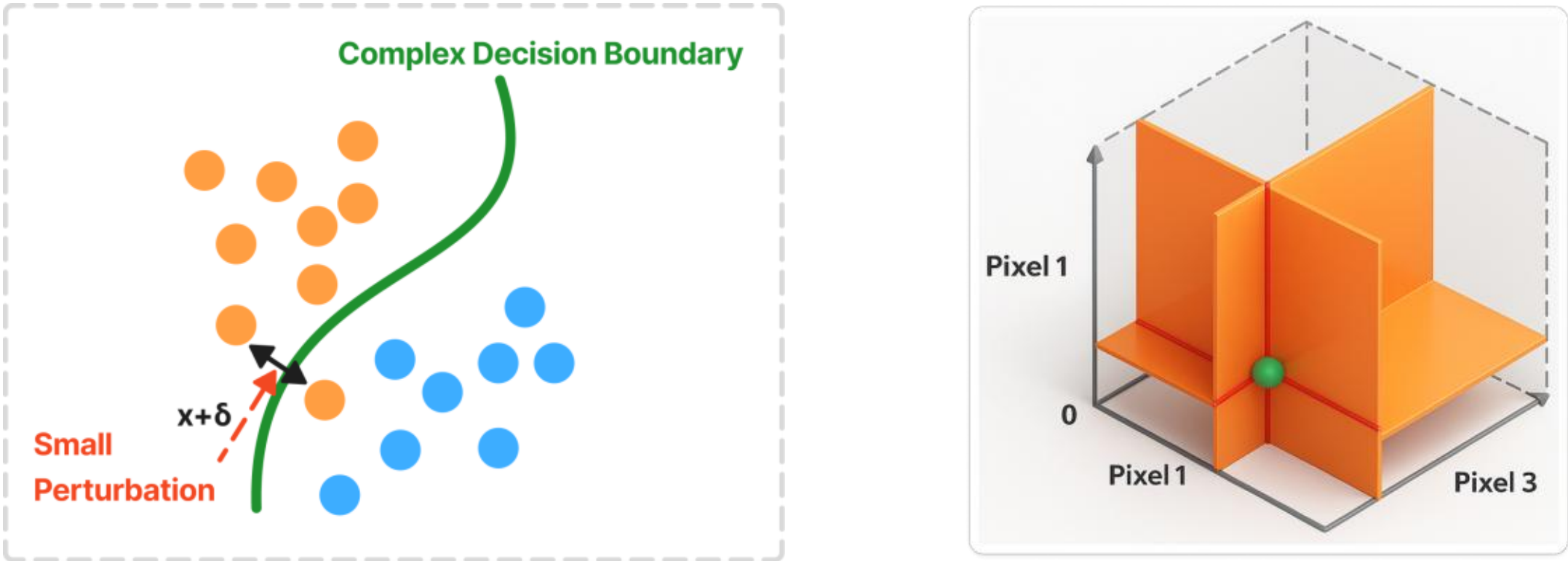

**(a)** **(b)**

**Figure 1.** (a) Illustration of single-pixel perturbation pushing input across non-linear decision boundary in feature space. (b) 3D visualization of one- and two-pixel perturbation attacks. The green sphere marks the original image. OPAs explore perturbations along three intersecting lines (1D), while two-pixel attacks span three shaded planes (2D) within the input space.

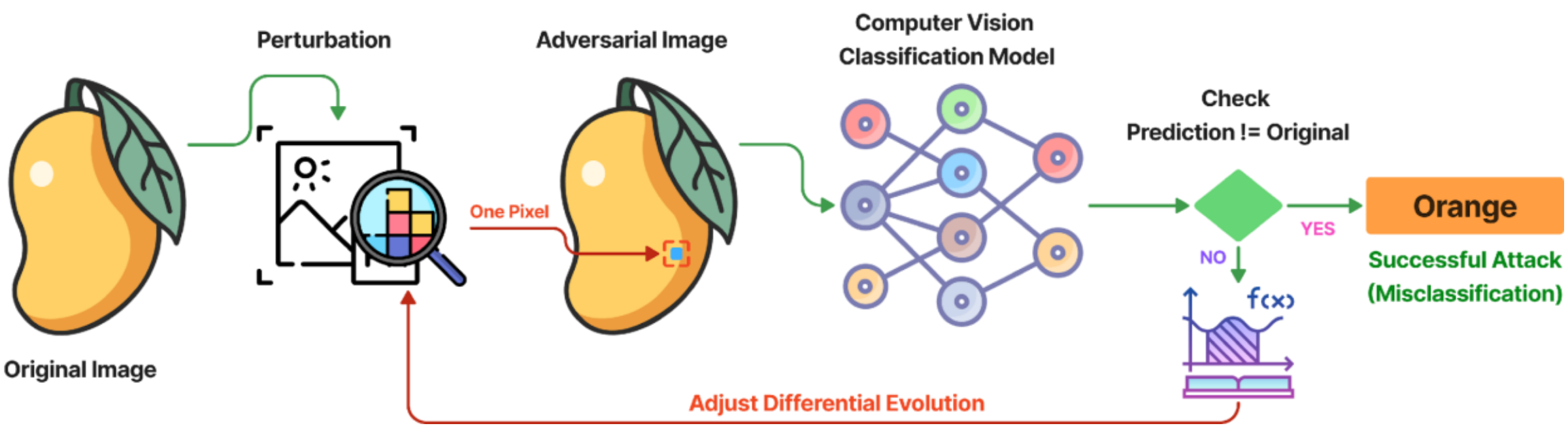

**Figure 2.** Visualization of OPA, where a minimal perturbation leads to misclassification by a computer vision model.

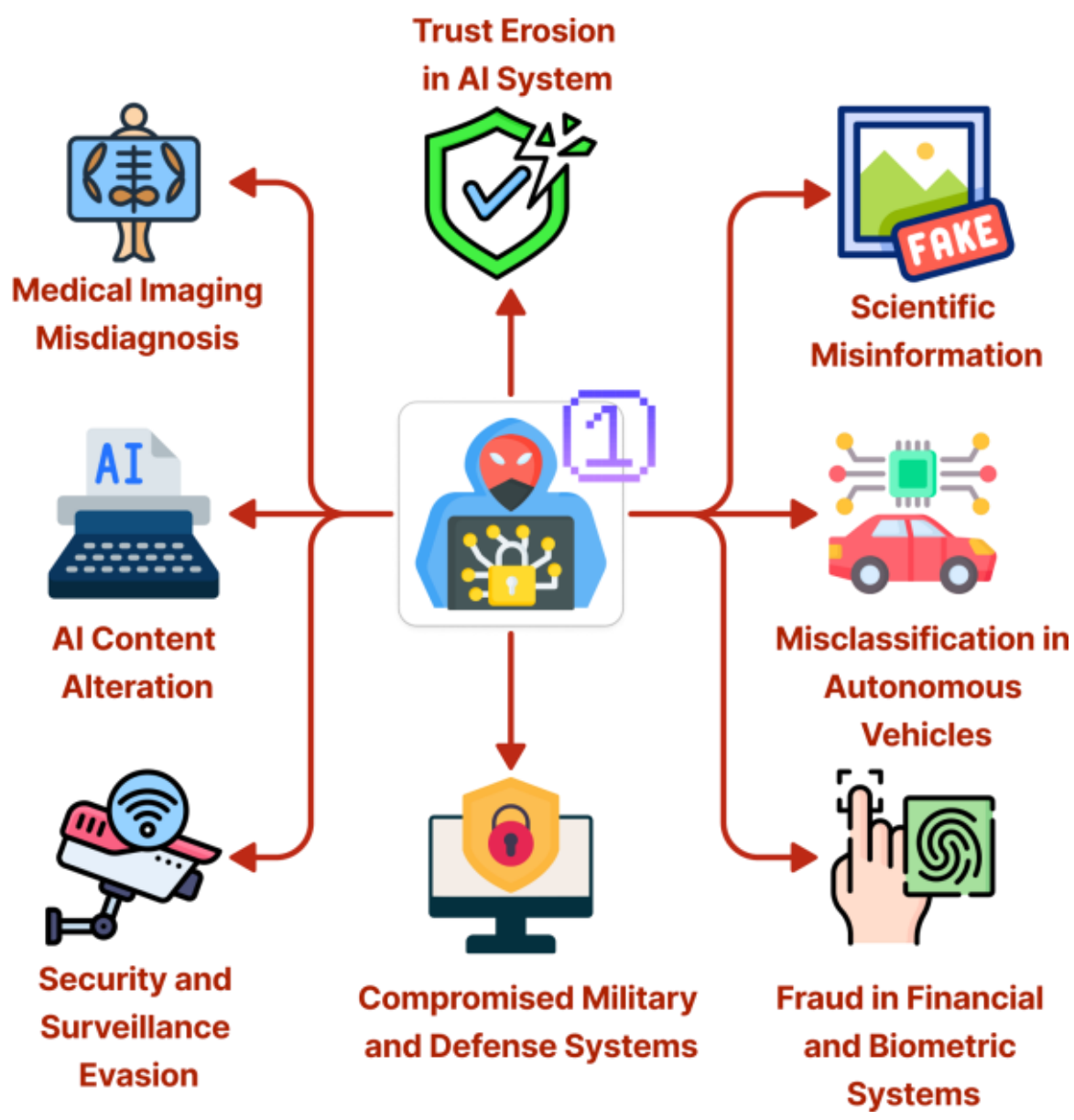

**Figure 3.** Potential impacts of OPAs across various domains, such as misclassification, evasion, trust erosion, and misinformation.

## 3. Review Methodology and Analytical Framework

This review follows a structured, PRISMA-guided methodology to ensure structured coverage and transparent synthesis of OPA literature published between 2017 and 2026. We applied a systematic search strategy, clear inclusion–exclusion criteria, and multi-stage screening to identify high-quality studies. Building on this, a multi-axis analytical framework was developed to classify the selected works across attack methods, model architectures, defence strategies, evaluation metrics, and domain applications. This compact but rigorous approach enables consistent comparison and highlights key patterns, limitations, and research gaps within the evolving OPA landscape. **Figure 4** presents the Graphical Abstract summarizing the OPA review workflow, from single-pixel perturbation effects and PRISMA screening to extracted insights, taxonomy, meta-analysis outcomes, and key contributions across methods, defenses, and applications.

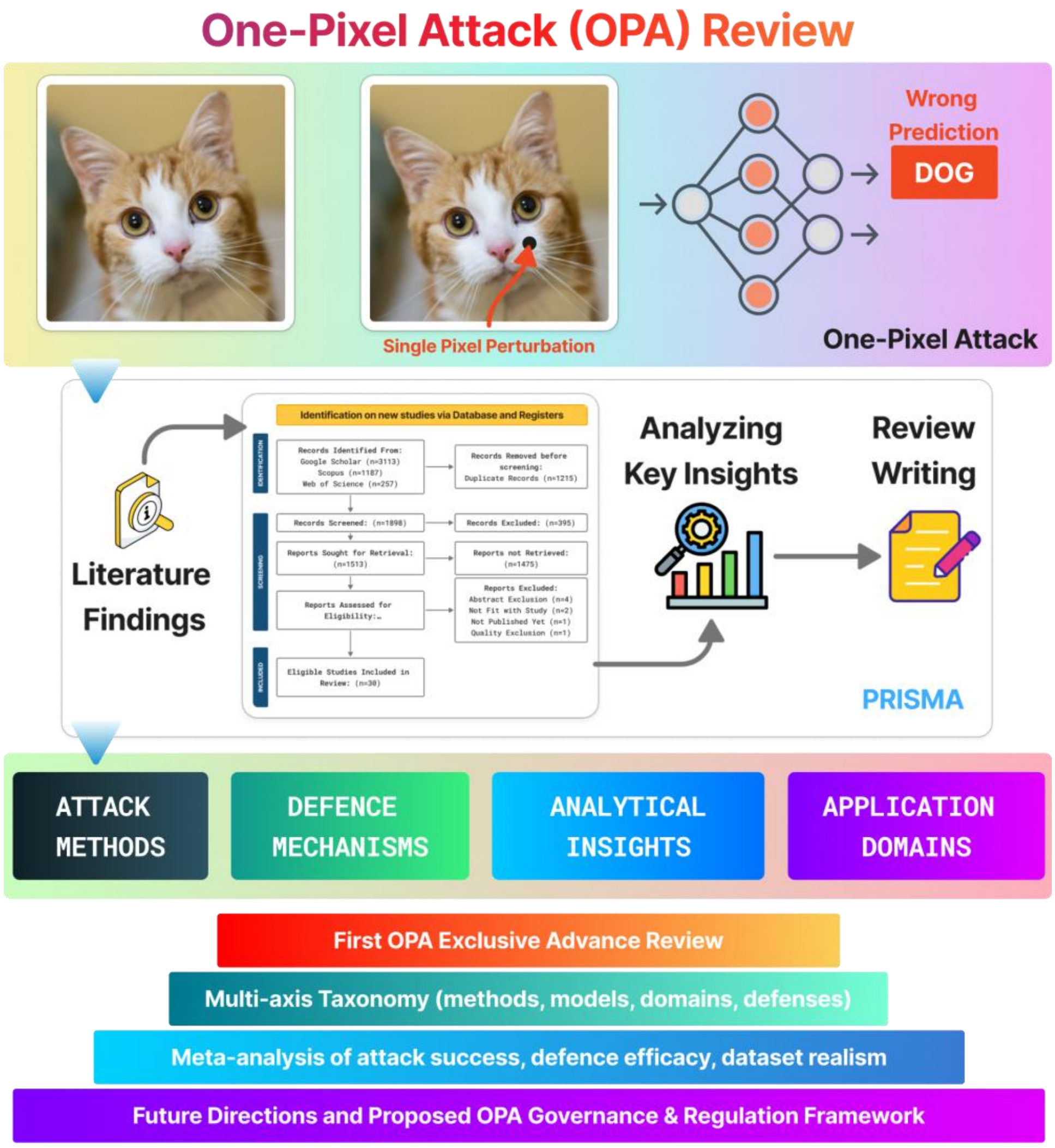

**Figure 4.** The PRISMA-guided systematic review mapping one-pixel adversarial attack methods, vulnerable domains, defence mechanisms, and policy recommendations across 32 studies published from 2017–2026.

### 3.1. Article Searching Strategy

To ensure systematic coverage of the literature on OPA in computer vision, a systematic and structured search strategy was employed in accordance with the Preferred Reporting Items for Systematic Reviews and Meta-Analyses (PRISMA 2020) guidelines [20]. The goal was to identify, screen, and select peer-reviewed studies that directly focus on the development, application, analysis, or defence of one-pixel or single-pixel adversarial attacks on deep learning models. Three widely recognized academic databases were selected based on their relevance, accessibility, and interdisciplinary coverage of computer vision, artificial intelligence, and adversarial machine learning: Scopus [21], Web of Science (WoS) [22], Google Scholar [23]. These databases were chosen to ensure inclusion of both highly cited and emerging literature, spanning journals, conferences, and preprints. The following Boolean search string was used to retrieve relevant articles: *("one-pixel attack" OR "single-pixel attack") AND ("computer vision" OR "adversarial attack" OR "deep learning" OR "image classification")*. This query was applied to titles, abstracts, and keywords to maximize coverage while minimizing irrelevant results. The search was restricted to the time frame between October 2017 and February 2026 to capture all significant advancements since the introduction of the original OPA by Su et al. [3] in October 2017. The inclusion of both "one-pixel attack" and "*single-pixel attack*" ensured linguistic variation was accounted for, while the secondary terms such as "*computer vision*," "*deep learning*," and "*image classification*" helped filter out irrelevant domains such as signal processing, robotics, or non-visual adversarial AI.

### 3.2. Inclusion and Exclusion Criteria

To ensure methodological rigor and maintain a focused scope aligned with the objectives of this review, a set of clearly defined inclusion and exclusion criteria was applied during the article selection process. These criteria guided the filtering of search results through title, abstract, and full-text assessments. Studies were included in the review if they met all of the following conditions:

a. The study directly addressed one-pixel or single-pixel adversarial attacks in computer vision tasks such as image classification, object recognition, or segmentation.
b. The attack was implemented on deep learning models including but not limited to CNNs, ResNet, VGG, DenseNet, or Vision Transformers.
c. Articles were published in reputable journals or conference proceedings, ensuring academic validity.
d. Publications between October 2017 and February 2026 were included to cover both foundational and the latest contributions.
e. Only articles written in English were considered.
f. The study included quantitative experiments, evaluations, or theoretical frameworks relevant to OPA or their defence mechanisms.
g. The paper provided sufficient description of models, datasets, attack/defence techniques, and evaluation metrics.

Articles were excluded from this review for any of the following reasons:

a. Studies that broadly discussed adversarial machine learning but did not mention or analyse OPA.
b. Papers centred on non-visual data such as speech, tabular, or time series attacks.
c. Articles lacking practical experiments or simulations related to OPA.
d. Conference papers later extended into journal versions were excluded to prevent double-counting.
e. These were removed during the eligibility check to maintain academic quality.
f. Articles that failed to provide information on dataset names, model configurations, or experimental results were deemed unfit for systematic review.

### 3.3. Article Selection Process

This review follows the PRISMA 2020 guidelines to ensure methodological transparency and replicability. The selection process was conducted in four stages: identification, screening, eligibility, and inclusion.

i. **Identification**: A comprehensive literature search was conducted across three major academic databases: Google Scholar, Scopus, and Web of Science (WoS). The search covered peer-reviewed publications from October 2017 to February 2026 using the query: ("one-pixel attack" OR "single-pixel attack") AND ("computer vision" OR "adversarial attack" OR "deep learning" OR "image classification"). The search retrieved a total of 3,929 records, including 3,113 from Google Scholar, 657 from Scopus, and 159 from Web of Science.
ii. **Screening**: Following the initial search, 3,647 duplicate records were removed using automated filtering tools, leaving 282 unique records for screening. These records were evaluated based on titles and abstracts. At this stage, 209 records were excluded due to lack of relevance, primarily because they did not focus specifically on pixel-level adversarial attacks or were outside the scope of computer vision. The remaining 73 records were considered potentially relevant.
iii. **Eligibility and Quality Assessment**: From the screened records, 73 full-text articles were successfully retrieved and assessed for eligibility. A total of 25 records could not be retrieved due to limited accessibility or insufficient availability. The retrieved 48 studies were evaluated using predefined inclusion criteria focusing on methodological rigor, clarity of objectives, and empirical validation. During this stage, 16 studies were excluded, including: 10 due to insufficient relevance based on abstract and full-text screening, 4 for not aligning with the study scope, 1 unpublished manuscript, and 1 study with poor methodological quality. A structured quality assessment framework was applied, emphasizing reproducibility, dataset diversity, and experimental robustness.
iv. **Inclusion**: After the eligibility assessment, a final set of 32 high-quality studies was selected for inclusion in the qualitative and quantitative synthesis. These studies form the foundation of this review and are analyzed in subsequent sections in accordance with the defined research objectives.

To support transparency, a graphical representation of this selection methodology is shown in **Figure 5**, and **Figure 6** presents the PRISMA flow diagram.

### 3.4. Critical Appraisal of the Literature

The 32 reviewed studies (2017–2026) reveal a field that is conceptually rich but methodologically uneven. Our structured coding of reproducibility, dataset diversity, and methodological rigor shows several clear patterns that shape the current state of OPA research. **Table 1** represents the summarized critical appraisal of our literature. By presenting publication details, research focus, citation metrics, and quality indicators such as reproducibility, dataset diversity, and methodological rigor, this table establishes both the breadth and credibility of the reviewed literature. The diversity across publishers and years reflects the evolving academic attention toward OPA, while the scoring framework clarifies which studies demonstrate stronger experimental soundness or broader dataset validation. Each reviewed study was critically appraised using three predefined methodological dimensions: reproducibility, dataset diversity, and methodological rigor. Reproducibility was evaluated based on the availability of implementation details, code repositories, and experimental transparency. Dataset diversity was assessed according to the breadth and heterogeneity of the datasets used, while methodological rigor was evaluated based on the clarity of the research objectives, experimental design, validation procedures, and analytical depth. Citation counts were considered only as contextual indicators of academic influence and were not used in the methodological appraisal. This structured qualitative assessment was used to identify methodological strengths, limitations, and research trends across the reviewed studies. The quantitative meta-analysis presented in **Figure 7** is derived from a systematic evaluation of 32 selected studies on one-pixel and ultra-sparse adversarial attacks. Each study was analyzed across four key dimensions: reproducibility, dataset diversity, methodological rigor, and citation impact. Reproducibility was assessed based on the availability of implementation details, code repositories, and experimental transparency, and categorized as either 'Yes' or 'Partial'. Dataset diversity was evaluated according to the breadth and heterogeneity of datasets used in each study, and classified into three levels: Low (single or highly constrained dataset), Moderate (multiple datasets within a domain), and High (cross-domain or large-scale diverse datasets). Methodological rigor was determined by examining the clarity of problem formulation, experimental design, validation procedures, and analytical depth, and categorized as Moderate or Strong. A composite quality score (1–5) was assigned to each study by synthesizing these dimensions. Studies with strong methodological foundations, higher dataset diversity, and clear reproducibility practices received higher scores, while those with limited validation or narrow datasets were assigned lower scores. Citation counts were included as an auxiliary metric to reflect academic influence but were not directly used in score computation.

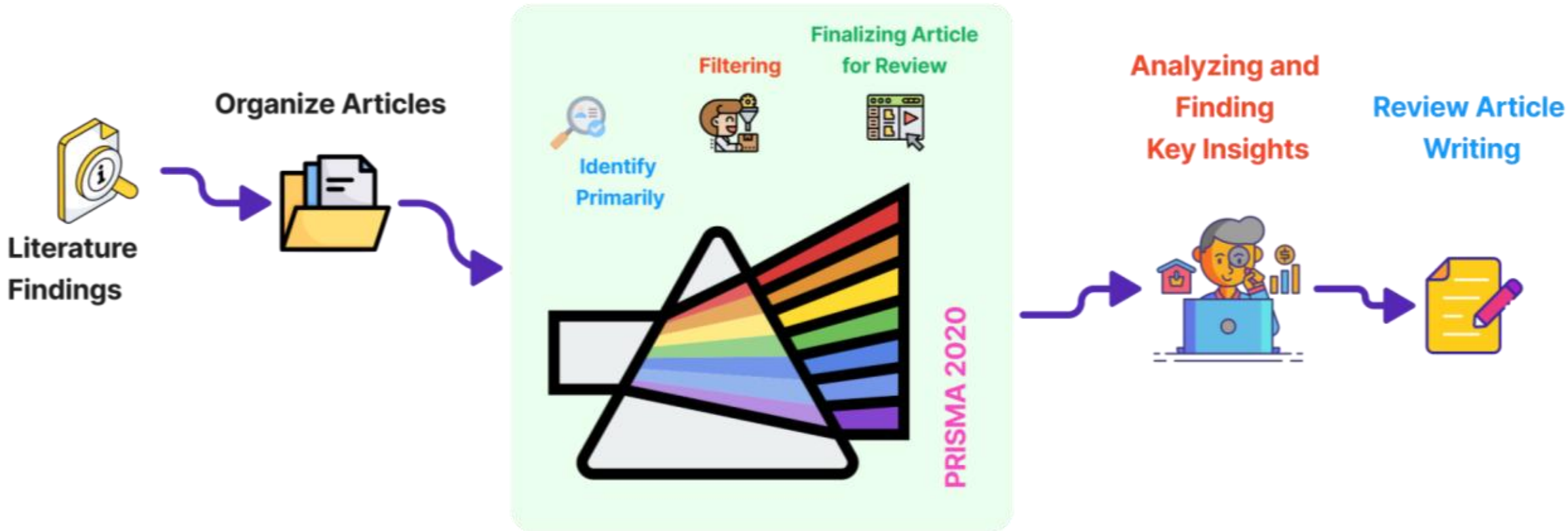

**Figure 5.** Overview of the review article writing process following the PRISMA 2020 framework.

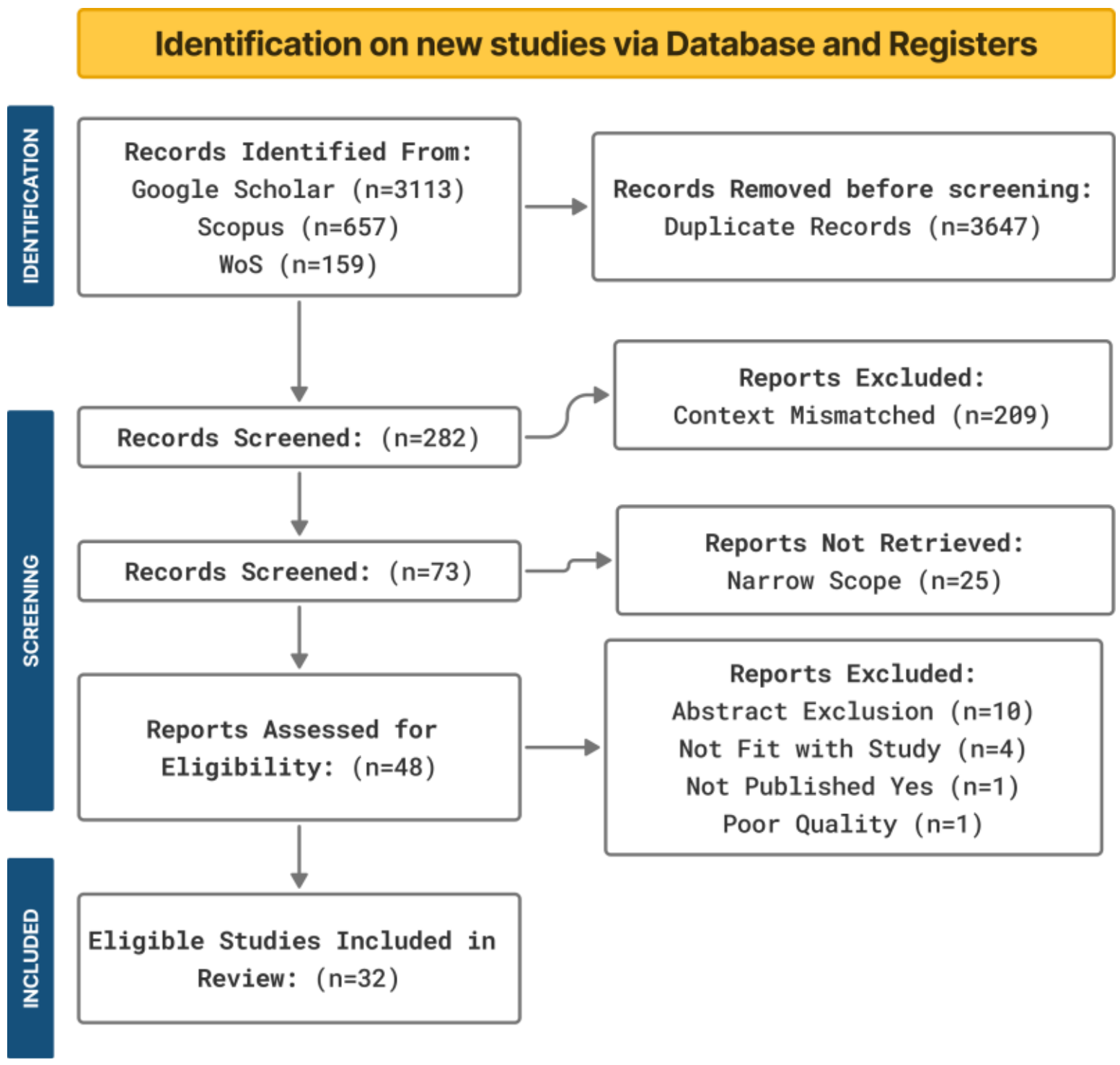

**Figure 6.** PRISMA 2020 flow diagram illustrating the study selection process for the systematic review

**Figure 7** presents a multi-dimensional visualization of the 32 studies included in the quantitative meta-analysis. Additional dimensions are encoded through visual attributes: bubble size represents dataset diversity, color indicates methodological rigor, and marker shape distinguishes reproducibility status. Several important trends emerge from this visualization. First, the majority of studies are clustered within the score range of 4 to 5, indicating that the field demonstrates generally strong methodological quality. However, despite this, most bubbles are relatively small, revealing that a significant proportion of studies rely on datasets with limited diversity. This suggests that while experimental design is often rigorous, the generalizability of findings may be constrained by narrow evaluation settings. Second, the relationship between citation count and study quality appears weak. Highly cited studies do not consistently correspond to the highest methodological scores, and several low-citation studies exhibit strong rigor and high scores. This observation highlights that citation impact alone is not a reliable indicator of research quality in this domain. Third, the predominance of circular markers indicates that most studies provide reproducible methodologies, although a small number of partially reproducible works remain. This reflects a positive trend toward transparency but also underscores the need for standardized reporting practices. Finally, studies utilizing higher dataset diversity (larger bubbles) tend to cluster at the upper end of the score spectrum, suggesting a positive association between dataset breadth and overall research quality. Nevertheless, such studies remain relatively few, reinforcing the observation that dataset diversity is a critical yet underdeveloped aspect of current research.

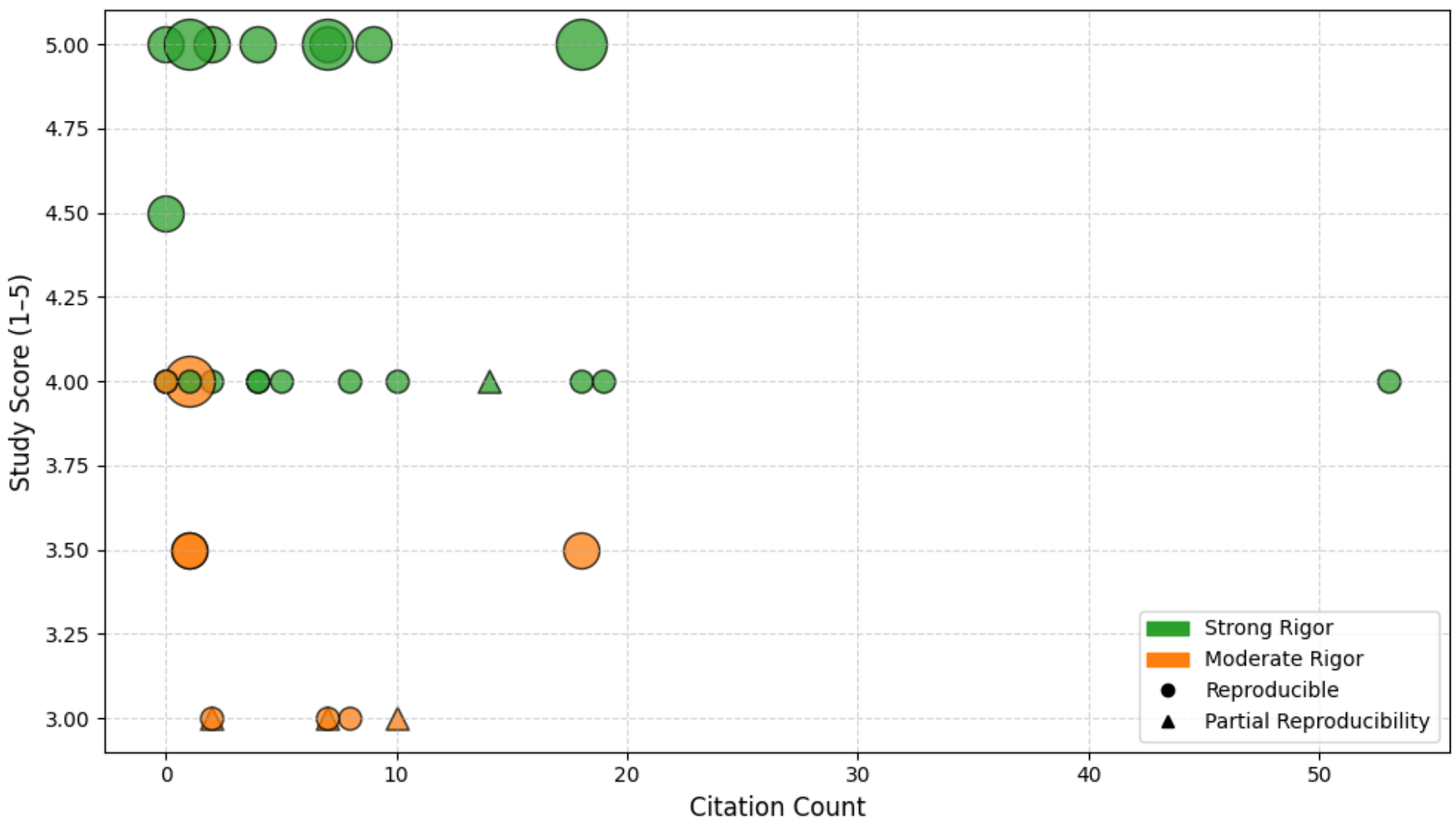

**Figure 7.** Multi-dimensional meta-analysis of all studies on one-pixel and ultra-sparse adversarial attacks. The x-axis represents citation count, and the y-axis denotes the assigned study quality score (1–5). Bubble size corresponds to dataset diversity (Low, Moderate, High), color indicates methodological rigor (Strong or Moderate), and marker shape represents reproducibility (circle: reproducible; triangle: partially reproducible). The visualization reveals that while most studies demonstrate strong methodological rigor, they predominantly rely on low-diversity datasets, and citation impact does not necessarily correlate with research quality.

Another important observation emerging from the critical appraisal is the architectural imbalance of the existing OPA literature. Although our literature search extended through February 2026, the empirical evidence remains heavily concentrated on conventional CNN architectures including LeNet, VGG, ResNet, DenseNet, AlexNet, and related variants. Only very limited work has investigated transformer-based vision models, and virtually no eligible studies provide standardized quantitative evaluations on Vision Transformers (ViTs), Swin Transformers, Masked Autoencoders (MAEs), or the vision encoders used in multimodal foundation models such as CLIP. Consequently, the current evidence base does not yet permit statistically rigorous comparison between CNNs and modern foundation architectures. This limitation should not be interpreted as evidence that newer architectures are inherently more robust or more vulnerable; rather, it reflects a significant gap in the published literature that warrants systematic future investigation.

**Table 1.** Structured Summary of Critical Appraisal

| Dimension | Key Findings | Evidence / Metrics | Implications |
|---|---|---|---|
| **Reproducibility & Transparency** | Majority of studies claim reproducibility but only few provide fully replicable code. | • 28/32 reproducible (87.5%)<br>• 4 partials (12.5%)<br>• 2 with full code repos | Validation remains difficult; reproducibility standards need strengthening. |
| **Dataset Diversity & Domain Spread** | Heavy dependence on MNIST, CIFAR-10, synthetic grids; limited real-world datasets. | • Majority "Low" diversity<br>• Only 4 "High" diversity studies | Limits generalizability; creates dataset bias;<br>real-world risks underexplored. |
| **Methodological Rigor** | Strong algorithmic rigor but narrow empirical scope; mostly controlled testbeds. | • Most rated "Strong" rigor<br>• Modern frontier architectures (Vision Transformers, Swin Transformers, MAEs, CLIP vision encoders and other foundation vision models) remain largely absent from current OPA empirical studies. | Current robustness conclusions are primarily supported by CNN-based evidence and cannot yet be generalized to contemporary transformer-based or multimodal architectures. |
| **Temporal Evolution** | Research shifted from toy benchmarks → domain-specific → refined evaluation. | • 2017–20: toy datasets, DE-driven<br>• 2021–22: domain expansion<br>• 2023–25: improved metrics | Field is maturing but unevenly; improvement in reporting and evaluation quality. |
| **Biases & Threats to Validity** | Systemic issues in benchmark choice, code release, and architecture coverage. | • Overuse of small datasets<br>• Limited ViT/large CNN testing<br>• Few complete pipelines | Hinders comparison across studies;<br>risks overestimating real-world vulnerability. |
| **Overall Synthesis & Required Shifts** | Strong conceptual progress, limited ecological validity; needs broader datasets and modern models. | • Identified four gaps: dataset breadth, model diversity, transparency, real-world context | Calls for standardized evaluations, domain-specific benchmarks, and open protocols. |

### 3.5. Research Questions and Categorizing Framework of OPA Literature

To ensure a structured and focused review, the following research questions were formulated. These questions and motivations guided the selection, classification, and evaluation of relevant studies and served to structure the synthesis of insights and identification of knowledge gaps represented in **Table 2.**

**Table 2.** Major Research Questions and Motivations of this Review

| **Questions** | **Motivations** | **Related Section** |
|---|---|---|
| **RQ1.** What are the dominant methodological approaches for constructing OPA, and how do their success rates and efficiency compare across diverse datasets and model architectures? | Despite the prevalence of differential evolution (DE) and bio-inspired algorithms (SBO, TLMFO, PSO) in current literature, comparative benchmarking is fragmented and dataset-specific. A comprehensive understanding of which optimization strategies are most effective and scalable is crucial for both attackers (seeking efficiency and higher misclassification) and defenders (identifying the highest-risk vulnerabilities). Illuminating the strengths and trade-offs of different optimization paradigms will drive both methodological innovation and robustness evaluation. | Section: 4, 7.2<br>Figure 6<br>Table 7 |
| **RQ2.** How effective are current defence mechanisms including denoising, autoencoder-based, transformation, and hybrid techniques in mitigating OPA, and what trade-offs do they introduce in practical scenarios? | While selective pixel denoising, VAE anomaly detection, autoencoder recovery, and data transformation (JPEG/compression, spatial smoothing) methods show promise, their efficacy varies widely depending on domain, image type, and computational constraints. Systematically quantifying recovery rates, detection abilities, and side effects (perceptual distortion, computational overhead) will inform the development of defences that are both practical and generalizable, especially in safety-critical applications such as medical imaging or industrial inspection. | Section: 5, 7.3<br>Table 8 |
| **RQ3.** How do OPA vulnerabilities and defence outcomes differ across focused application domains, such as medical imaging, quantum communication, and industrial systems? | Domain-specific case studies highlight unique challenges such as a heightened risk of misdiagnosis in pathology, security breaches in quantum key distribution systems, or integrity threats in traffic sign recognition. Understanding domain-driven differences in attack/defence performance will reveal where AI safety measures are most urgently needed, guide the design of tailored solutions, and inform regulatory bodies regarding the deployment of computer vision across critical sectors. | Section: 6, 7.4<br>Table 9 |
| **RQ4.** What are the principal factors influencing the imperceptibility and detectability of OPA, especially concerning model | The stealthiness of OPA remains controversial, with evidence that attacks often align with high-saliency regions or model receptive field boundaries, while their visual imperceptibility may not always hold up in real environments. Analysing how architectural features (residual connections, input resolution) and image properties impact | Section: 4, 7.1<br>Figure 6 |
| structure and input data characteristics? | both attack success and defence/detection accuracy will provide actionable insights for designing inherently robust systems. | |
| **RQ5**. What research gaps persist regarding the transferability, scalability, and real-world robustness of OPA and their defences, and how should future work address them? | Most studies are limited to controlled settings (small datasets, simulated attacks). There are substantial open questions about model-to-model and virtual-to-physical transferability, the impact of increasing pixel budgets, and the scalability of proposed solutions to large-scale or high-resolution data. Identifying and confronting these limitations will help establish standardized benchmarks, guide the research agenda toward explainability and domain generalization, and promote trustworthy deployment of AI systems. | Section: 7.5 |

To ensure conceptual coherence across heterogeneous OPA studies, we developed a four-axis classification framework illustrated in **Figure 8**.

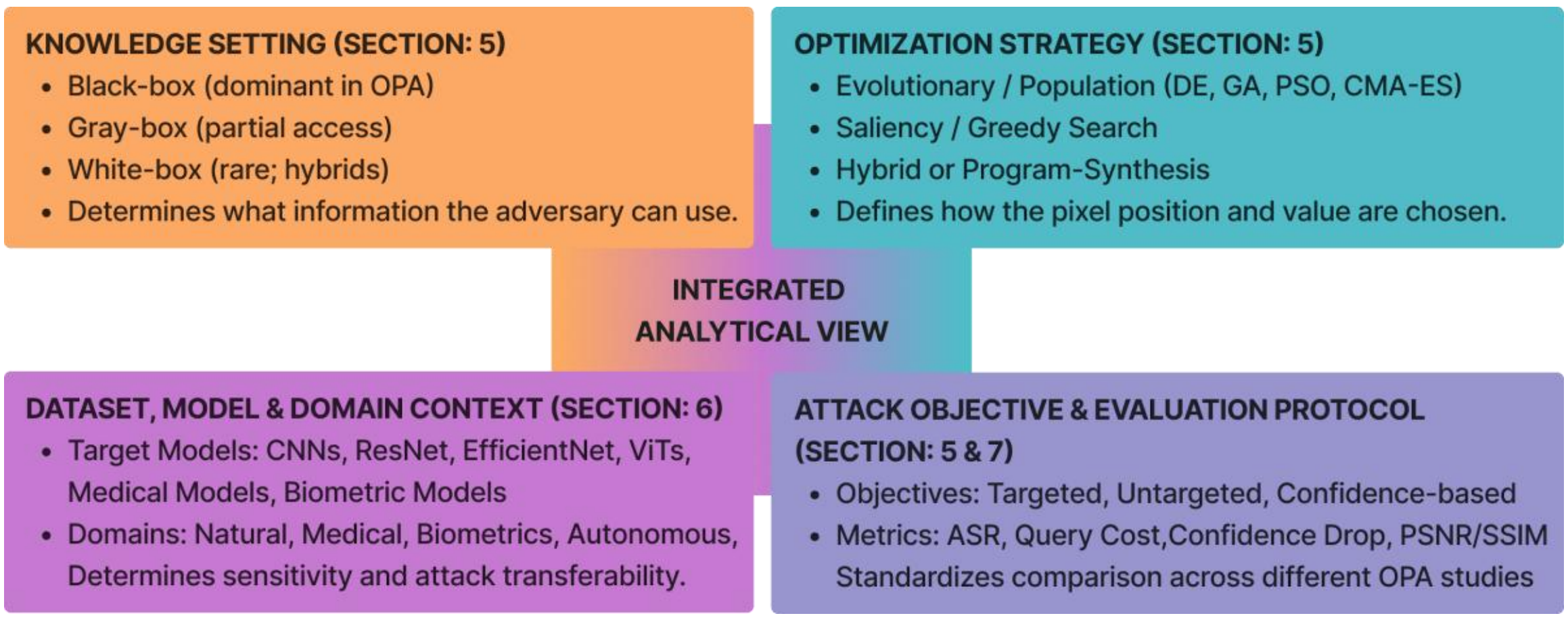

**Figure 8.** Different Axis of Analytical Framework for Categorizing OPA Literature

**Table 3** complements this by standardizing terminology across the field. The extensive list of acronyms and their full forms ensure clarity in interpreting diverse algorithms, optimizers, and network architectures used throughout the review. Beyond functioning as a glossary, it underscores the interdisciplinary nature of OPA research drawing from computer vision, optimization theory, and security analytics.

**Table 3.** Acronyms and Abbreviations

| **Acronym** | **Full Form** |
|---|---|
| CNN [24] | Convolutional Neural Network |
| ResNet [25] | Residual Neural Network |
| VGG16/ VGG19 [26] | Visual Geometry Group (16/19-layer) |
| LeNet [27] | LeNet-5 |
| NiN [28] | Network in Network |
| DenseNet [29] | Densely Connected CNN |
| GoogLeNet [30] | Google Inception Network |
| AlexNet [31] | AlexNet |
| WideResNet [32] | Wide Residual Network |
| HybridSN [33] | Hybrid Spectral-Network |
| FER-net [34] | Facial Expression Recognition Network |
| DE [13] | Differential Evolution |
| FGSM [35] | Fast Gradient Sign Method |
| PGD [36] | Projected Gradient Descent |
| JSMA [37] | Jacobian-based Saliency Map Attack |
| C&W [38] | Carlini & Wagner Attack |
| BIM [39] | Basic Iterative Method |
| I-FGSM [40] | Iterative Fast Gradient Sign Method |
| MI-FGSM [41] | Momentum Iterative FGSM |
| L-BFGS [42] | Limited-memory Broyden–Fletcher–Goldfarb–Shanno |
| SBO [43] | Smell Bees Optimization |
| OPPSLA [44] | One-Pixel Program Synthesis via Metropolis-Hastings |
| Sparse-RS [45] | Sparse Random Search |
| SuOPA [46] | Surrogate-based One-Pixel Attack |
| TLMFO [47] | Teaching-Learning-based Moth-Flame Optimizer |
| MFO [48] | Moth-Flame Optimization |
| GBO [49] | Gradient-Based Optimizer |
| AOA [50] | Arithmetic Optimization Algorithm |
| SOA [51] | Seagull Optimization Algorithm |
| RSA [52] | Reptile Search Algorithm |
| RISOPA [53] | Rapid Imperceptible Strong OPA |
| OPA2D [54][122] | One-Pixel Attack with Detection/Defence |
| IOPA [55] | Improved One-Pixel Attack |
| CMA-ES [56] | Covariance Matrix Adaptation Evolution Strategy |
| PSO [57] | Particle Swarm Optimization |
| SGD [58] | Stochastic Gradient Descent |
| MSE [59] | Mean Squared Error |
| ROC [60] | Receiver Operating Characteristic |
| AUC [61] | Area Under the Curve |
| SLIC [62] | Simple Linear Iterative Clustering |
| PSNR [18] | Peak Signal-to-Noise Ratio |
| SSIM [19] | Structural Similarity Index Measure |

## 4. Taxonomy and Mapping of the OPA Landscape

The taxonomy in **Figure 9** illustrates the structural framework of OPA research, showing how attack design, model architecture, and defence strategies interact across multiple analytical layers. At the core, black-box evolutionary algorithms such as DE, SBO, and PSO dominate, using stochastic population-based searches to locate the minimal pixel perturbation that maximizes misclassification. Hybrid meta-heuristic extensions enhance convergence speed and stealth, while gradient-based methods serve as white-box baselines. The model axis reveals that lightweight CNNs and LeNet are highly sensitive to localized perturbations, whereas deeper residual or dense networks show partial resilience through skip connections and distributed gradient flow. Defence mechanisms span pixel-level restoration (Noise2Noise), input-space transformation (JPEG compression), and architectural hardening (fractional-order adversarial training, autoencoders), reflecting the field's balance between image fidelity and robustness. Supporting techniques saliency and activation maps, SLIC super-pixels, Latin Hypercube Sampling highlight the growing role of interpretability-guided analysis in mapping vulnerability regions and validating attacks. Performance evaluation in these studies combines ASR, confidence drop, and imperceptibility metrics such as PSNR, SSIM, and $L_2$ norms to gauge both effectiveness and visual subtlety. Efficiency is captured through query count and runtime per image, where optimized variants like RISOPA achieve up to 57× speedups. Defence assessment relies on accuracy restoration, detection precision, and AUC-ROC, indicating recovery strength and generalizability. Together, these elements form a unified view of OPA research linking algorithmic innovation, architectural vulnerability, and quantitative performance into a coherent taxonomy of adversarial behaviour in deep learning.

**Table 4** serves as a consolidated map of the experimental ecosystem underlying all reviewed OPA studies. By aligning attack models, optimization strategies, and defence mechanisms within a single structured frame, it clarifies how different algorithmic families and architectures intersect. This synthesis allows direct comparison between evolutionary, gradient-free, and hybrid attack approaches, and highlights which neural networks such as ResNet, VGG, or DenseNet are most frequently targeted. It also illustrates how supporting techniques like saliency mapping, super-pixel segmentation, and statistical testing are embedded within OPA workflows.

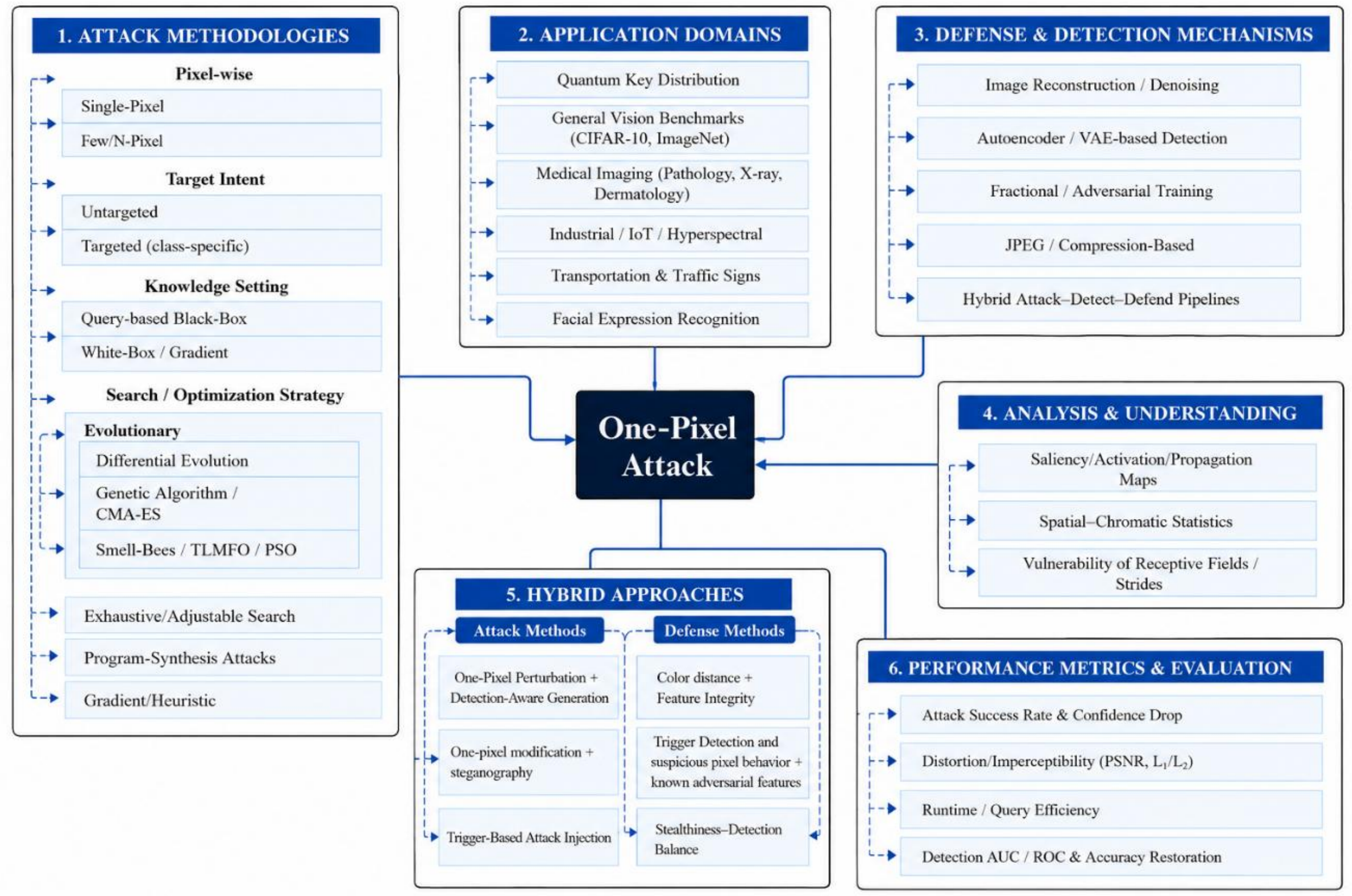

**Figure 9.** Taxonomy of OPAs, covering methodologies, metrics, applications, and evaluation dimensions.

**Table 4.** Categorization of models, methods, defences, attacks, and optimizers used in all OPAs' studies

| Generic Category | Subcategory | Methods & Models | Article IDs |
|---|---|---|---|
| **Model Architectures** | DNN | Raw CNN | [63,64,65,41,66,67,69,68,70,71,53] |
| | | ResNet | [72,3,65,73,74,75,76,77,78,79,80,81,53,82] |
| | | VGG16 | [83,68,80,84,81] |
| | | VGG19 | [84] |
| | | LeNet | [3,73,77,78,80,84,53] |
| | | NiN | [63,72,65,73,41,77,68,78,79,80,70,55,84,81,53] |
| | | DenseNet | [78,79,84,53,85] |
| | | GoogLeNet | [79] |
| | | AlexNet | [80] |
| | | WideResNet | [53] |
| | | PureCNN | [53] |
| | | DeepVGG16 | [86] |
| | | HybridSN | [55] |
| | | TrafficSignNet | [67] |
| | | IBM MAX (ResNet-50) | [75,87] |
| **Adversarial Attacks** | Gradient-based | FGSM | [74,41,83,55,84] |
| | | PGD | [4,55,84] |
| | | BIM | [41,55] |
| | | I-FGSM | [41] |
| | | MI-FGSM | [41] |
| | | JSMA | [72,64,41,78] |
| | | Carlini & Wagner | [41,70] |
| | | DeepFool | [84] |
| | | L-BFGS | [41] |
| | Optimization/Evolutionary | Differential Evolution | [63,3,4,74,75,41,54,88,66,67,76,83,68,79,70,71,81,85,82] |
| | | Smell Bees Optimization | [43][86] |
| | | Particle Swarm Optimization (PSO) | [81] |
| | | AOA | [80] |
| | | GBO | [80] |
| | | CMA-ES | [70] |
| | | SOA | [80] |
| | | SuOPA | [78] |
| | | MFO | [80] |
| | | TLMFO | [80] |
| | Specialized/ Targeted | Single-Pixel Attack | [3,41,77] |
| | | Improved One-Pixel Attack | [55] |
| | | Sparse-RS | [68] |
| | | OPPSLA | [78] |
| | | OPA2D-ATK | [54] |
| | | OPA2D-DET | [54] |
| | | CornerSearch | [78] |
| | | Pointwise | [78] |
| | | RISOPA | [53] |
| | | RSA | [80] |
| **Défense Mechanisms** | Noise & Filtering | Noise2Noise | [72] |
| | | JPEG Compression | [83] |
| | Robust Training & Augmentation | Data Augmentation | [68,84] |
| | | Feature Squeezing | [80] |
| | | Interval Bound Propagation (IBP) | [84] |
| | | Fractional-Order Adversarial Training | [55] |
| | | Convolutional Autoencoder | [84] |
| | Regularization/Detection | Gradient Masking | [4] |
| | | Robust Optimization | [4] |
| **Supporting & Analytical Methods** | Explainability & Visualization | Saliency Maps | [64,3] |
| | | Activation Maps | [74] |
| | | Propagation Maps | [124] |
| | | Pixel Adversarial Maps | [64] |
| | Sampling & Segmentation | SLIC Superpixel Segmentation | [81] |
| | | Latin Hypercube Sampling | [66] |
| | Search & Computation | Exhaustive Search | [63,69] |
| | | GPU-based Parallel Processing | [69] |
| | Evaluation & Testing | Minkowski Distance | [87] |
| | | Wilcoxon Tests | [80] |
| **Optimizers** | Core Training Optimizers | Adam | [63,87,81] |
| | | SGD | [55] |

**Table 5.** Target Architectures and Algorithmic Attack Strategies Synthesized

| **Attack Strategy / Algorithm** | **Most Targeted Models (Architectures)** | **Common Result Metrics Tracked** | **Highlighted Performance Outcome** |
|---|---|---|---|
| Differential Evolution (DE) | Custom CNNs, VGG-16 | Attack Success Rate (ASR) | Highly successful on low-res, computationally expensive for high-res. |
| Momentum Finite-Difference (OPA4HR) | ResNet-50, DenseNet-121, VGG-16 | ASR, Query Efficiency, Distance | ~50.3% reduction in query cost; improved score-based settings. |
| One-Pixel Blackout / torchattacks | ResNet-50 (Medical Classifiers) | Accuracy Drop, Uncertainty Score | Accuracy drop (e.g., 88% → 69%); reveals high-risk overconfident regions. |

While **Table 4** categorizes the extensive range of architectures and optimizers utilized across the reviewed literature, it is equally important to synthesize how these elements interact in practice. **Table 5** provides a consolidated overview of the most prominent algorithmic attack strategies mapped against their primary target architectures. Across the reviewed studies, Differential Evolution (DE) appears most frequently as the baseline optimization strategy. However, because the studies employ heterogeneous datasets, model architectures, and evaluation protocols, its reported performance should be interpreted as evidence of widespread adoption rather than definitive algorithmic superiority.

## 5. Evolution and Diversification of OPA Methodologies

OPA represent a critical class of adversarial techniques that reveal the extreme sensitivity of DNN to even the smallest perturbations [91]. First introduced by Su et al. [77], the central premise of OPA is that altering a single pixel in an image can cause significant misclassification, despite the change being imperceptible to human observers. This attack operates in a black-box setting [90], requiring only output probabilities rather than access to internal model parameters, and typically employs Differential Evolution (DE) for perturbation optimization. Early empirical results highlighted the attack's effectiveness, with success rates reaching up to 68% on CIFAR-10, but dropping to around 16% on ImageNet due to increased input dimensionality and richer feature representations [75]. These findings underscored the role of input complexity and dataset structure in adversarial robustness. Moreover, theoretical studies revealed that OPA often target high-saliency regions [92], which are crucial to a model's decision-making process yet visually subtle. This alignment between adversarial impact and model attention maps has informed the design of both attacks and corresponding defences [89]. Analyses of receptive field vulnerabilities [93], decision boundary proximity [94][123], and pixel-level attribution maps have further emphasized the fragility of DNNs to localized perturbations [95]. The evolution of OPA methodologies can be broadly grouped into six emerging trends: (1) Optimization-based attacks, (2) Hybrid/metaheuristic approaches, (3) Program synthesis-based methods, (4) Physically realizable attacks, (5) Efficiency-optimized designs, and (6) multi-functional attack frameworks.

Early works predominantly relied on evolutionary optimization [96, 97]. DE quickly became a cornerstone, achieving up to 42% success on CIFAR-10/ImageNet [65] through iterative pixel refinement. Alternatives such as Genetic Algorithms (GA) improved pixel coverage but lagged in efficiency, while Covariance Matrix Adaptation Evolution Strategy (CMA-ES) offered faster convergence for small-scale datasets like GTSRB [65]. Nevertheless, these methods faced drawbacks such as slow convergence in high-dimensional spaces [8] and significant computational overhead [116]. To address these limitations, researchers developed hybrid methods that combine evolutionary logic with bio-inspired heuristics. For instance, Smell Bee Optimization (SBO) [22] achieved 80% success in localized attacks on MNIST and CIFAR-10 by focusing on semantically critical edge regions. Similarly, Teaching-Learning-based Moth-Flame Optimizer (TLMFO) [3] demonstrated statistically superior robustness ($p < 0.05$) for industrial defect detection, while Particle Swarm Optimization (PSO) [41] halved the computational time in facial recognition tasks by optimizing within super pixel domains. A major leap occurred with OPPSLA [72], which employed program synthesis using Metropolis-Hastings search to generate adversarial "programs" rather than direct pixel modifications. This approach required three times fewer queries than DE and improved cross-model transferability (from VGG-16 to ResNet18), making it particularly suitable for query-limited black-box environments. However, its success came with the risk of overfitting to pre-defined condition sketches. The real-world applicability of OPAs was explored by Liu et al. [18], who demonstrated physical attacks against traffic sign recognition systems. Their experiments showed 25% success at 0°–45° viewing angles, although environmental distortions such as lighting reduced classifier confidence by 9.2–10.9%, signalling serious concerns for autonomous vehicle safety. Subsequent innovations focused on enhancing efficiency and practicality. CriPI [98] achieved 83% success on CIFAR-10 while reducing attack generation time to just 1 minute per image, significantly outperforming DE in runtime (7× faster). Similarly, RISOPA [53] combined evolutionary search with gradient-inspired heuristics, achieving 57× speedups while maintaining competitive attack success rates, thus enabling real-time adversarial testing in sensitive applications. Beyond simple misclassification, OPAs have evolved into multi-functional tools. Some studies embedded covert information within single-pixel perturbations up to 11 bits per image at 30–50 dB PSNR demonstrating their potential in data hiding and steganography [18]. The OPA2D-ATK framework [21] unified attack, detection [122], and recovery into a single pipeline, reducing color distance by 20% while preserving imperceptibility, thus expanding adversarial utility into broader security and forensic applications. **Figure 10** illustrates the evolution of OPA strategies over time. These advancements illustrate a clear evolutionary trajectory: from brute-force search to informed and adaptive heuristics that enhance success and reduce computational cost. The shift reflects growing maturity in the field, aligning sparse perturbation strategies with biologically inspired optimization and domain-specific intelligence.

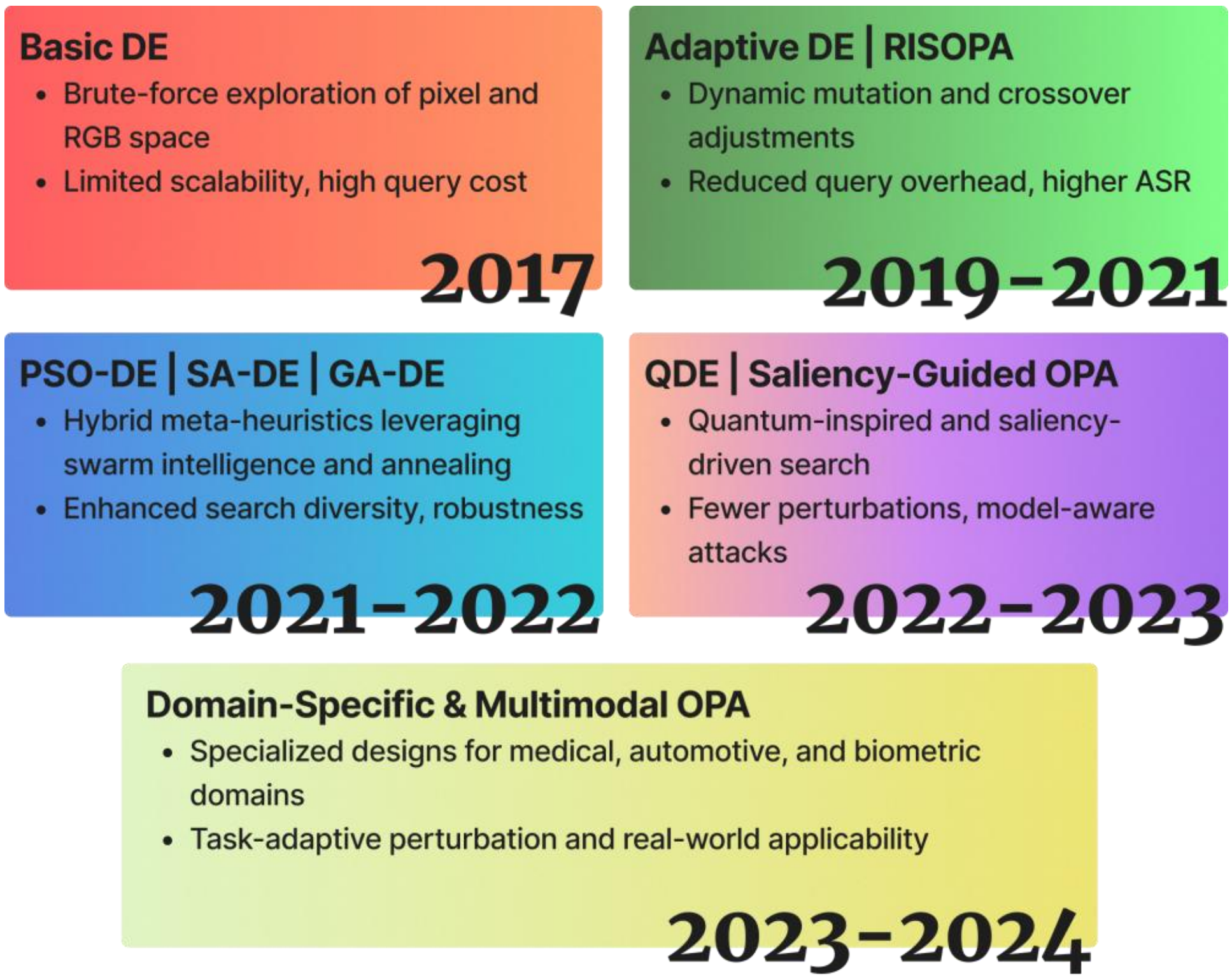

**Figure 10.** Evolution of OPA strategies over time

The trajectory of OPA research reveals a distinct shift in academic focus, adapting to the increasing complexity of target models and the demand for query-efficient attacks. As illustrated in **Figure 10**, the field has transitioned from foundational brute-force methods to domain-specific heuristics. To further contextualize this progression, **Table 6** outlines the temporal distribution of OPA literature from 2018 to 2026. This summary highlights how the primary problem statements have evolved—moving from establishing baselines on synthetic datasets to addressing the severe computational bottlenecks associated with high-resolution inputs and the complex decision boundaries of medical AI.

**Table 6.** Temporal Distribution and Focus Areas of OPA Literature

| Year Range | Primary Focus / Problem Statements | Common Datasets Used | Key Obstacles Addressed |
|---|---|---|---|
| **2018–2020** | Vulnerability of CNNs, establishing baselines for minimal perturbations. | Synthetic data, CIFAR-10 | Black-box interpretation, optimization failures, exhaustive search costs. |
| **2021–2023** | Defenses, defect detection, and backdoor clean-label attacks. | Standard RGB, ImageNet | Off-manifold behavior, lack of clear decision boundaries. |
| **2024–2026** | High-resolution inputs, medical AI fragility, query-efficient algorithms (e.g., OPA4HR). | BUSI (Clinical Ultrasound), High-Res RGB | High query costs in black-box settings, medical image complexity, inefficiency of DE-methods. |

**Table 7** provides a comparative summary of key studies focusing on the theoretical and empirical analysis of OPA. We explore the evolutionary progression of methodologies, spanning six primary trends: optimization-based approaches such as Differential Evolution, hybrid or metaheuristic techniques, program synthesis methods, physically realizable attacks, efficiency-optimized strategies, and multi-functional designs. Then we apply a functional framework evaluating Algorithmic Foundations, Performance Metrics, and Practical Utility, as detailed in **Table 8**.

**Table 7.** Key insights on attack success rates, targeted domains, and implications for defence strategies

| Article ID | Domain | Key Challenge | Dataset (Size) | Attack Success Rate | Critical Insight | Implications for Defence |
|---|---|---|---|---|---|---|
| [64] | General CV | Attack position vs. saliency maps | CIFAR-10 (800) | 79.3% (high-saliency) | Attacks target semantically meaningful regions (edges, object boundaries). | Defences must prioritize high-saliency areas (patch denoising). |
| [3] | General CV | Receptive field vulnerability | CIFAR-10 (5,000) | 58.68% (LeNet) | Simpler models (LeNet) are more | Architectural hardening |
| | | | | | vulnerable due to coarse receptive fields. | (residual connections) reduces susceptibility. |
| [74] | General CV | Imperceptibility vs. robustness | CIFAR-10/ImageNet (60,000) | 63–71% (CIFAR-10) | Low-resolution datasets (CIFAR) are easier to perturb than high-resolution datasets (ImageNet). | Input dimensionality impacts robustness; up-sampling may help. |
| [75,76] | Medical Imaging | Stealth requirements | TUPAC16 (4,620) | 91% (mitosis→normal) | Grayscale medical images are vulnerable to subtle perturbations in diagnostic regions. | Domain-specific defences (variational autoencoders) are needed for clinical safety. |
| [54] | General CV | Detectability trade-offs | CIFAR-10 (1,000) | 49.41% (OPA2D detection) | Human-imperceptible attacks evade traditional detection (JPEG). | Hybrid detection (OPA2D's attack+ defence) improves robustness. |
| [88] | General CV | White-box assumptions | CIFAR-10 (60,000) | 30.1% (C=5 candidates) | Detection fails without model internals (black-box limitation). | Anomaly detection (VAEs) works better in black-box settings. |
| [77] | Data Hiding | Stealth vs. payload capacity | CIFAR-10 (60,000) | 11 bits/image (PSNR 30–50 dB) | Single-pixel modifications enable covert communication. | Adversarial training disrupts hidden payloads without degrading clean data. |

It is important to interpret the comparisons presented in **Tables 7** and **Table 8** within the context of the underlying experimental heterogeneity. The reviewed studies employed different datasets, target architectures, attack objectives, optimization settings, and evaluation protocols. Consequently, reported attack success rates cannot be interpreted as directly comparable indicators of algorithmic superiority. Instead, the comparative synthesis is intended to identify recurring methodological characteristics, practical strengths, computational trade-offs, and application-specific performance patterns across the literature. A statistically rigorous ranking of optimization algorithms would require standardized evaluation under identical benchmark datasets, target models, and experimental protocols, which is currently lacking in the OPA literature.

**Table 8**. Methodological comparison of OPA across various datasets

| Category | Attack Method | Key Features | Strengths | Limitations | Success Rate | Models Tested |
|---|---|---|---|---|---|---|
| | Differential Evolution (DE) [**63**] | - Black-box optimization.<br>- Iterative pixel search. | - High success rates (42–71%).<br>- Model-agnostic. | - Slow convergence.<br>- High computational cost. | 42–71% (CIFAR-10, ImageNet) | ResNet, LeNet, CNN |
| **Evolutionary Algorithms** | Genetic Algorithms (GA) [**3**] | - Pixel coverage optimization.<br>- multi-generation refinement. | - Balanced distortion/success. | - Less efficient than DE. | ~40% (CIFAR-10) | Distilled CNNs |
| | Covariance Matrix Adaptation (CMA-ES) [70] | - Fast convergence.<br>- Adaptive parameter tuning. | - Efficient for small datasets. | - Low diversity in solutions. | 26–50% (GTSRB, MNIST) | TrafficSignNet, CNN |
| **Gradient-Free Optimization** | Smell Bees Optimization (SBO) [86] | - Mimics a bee foraging.<br>- Focuses on critical regions. | - 80% success in targeted regions (R2, R4). | - Less effective on RGB images. | 24–82.8% (MNIST/CIFAR-10) | ResNet, DeepVGG16 |
| | Particle Swarm Optimization (PSO) [81] | - Superpixel-focused (facial expressions).<br>- Faster than DE. | - 77% success (JAFFE).<br>- Reduced runtime. | - Fails outside critical regions. | 26–77% (FER datasets) | VGG16, ResNet50 |
| **Program Synthesis** | OPPSLA (Metropolis-Hastings) [78] | - Generates adversarial programs.<br>- Minimal queries. | - 3× faster than baselines.<br>- High transferability. | - Overfitting risk in condition language. | High (CIFAR-10/ImageNet) | VGG-16, ResNet18 |
| **Hybrid Methods** | CriPI (Critical Pixel Iterative) [98] | - Identifies critical pixels iteratively.<br>- Balance's speed/success. | - 83% success (5-pixel).<br>- 1 min/image. | - Limited to low-resolution images. | 61–86% (CIFAR-10) | CNN (CIFAR-10) |
| | RISOPA (Gradient Ascent + Momentum) [53] | - Combines gradient hints with evolutionary search.<br>- 57× faster than DE. | - Imperceptible perturbations.<br>- Scalable. | - Requires partial model access. | Comparable to DE (CIFAR-10) | ResNet, DenseNet |
| **Brute-Force** | TLMFO (Teaching-Learning MFO) [80] | - Hybrid of MFO and DE.<br>- Targets ZDM security. | - Superior to DE/MFO ($p < 0.05$).<br>- Robust. | - Complex implementation. | High (ImageNet) | AlexNet, VGG16 |

**6. Overview of Domain-Specific OPA**

OPA pose significant threats to high-stakes domains, particularly in medical [99][100][101] and quantum systems [102][103][104], where precision is critical. The conceptual framework in a study [4] reveals that even minimal adversarial perturbations can compromise AI-assisted medical diagnoses, raising both ethical and clinical concerns. This vulnerability is further evidenced in [66], where a one-pixel alteration was sufficient to alter the result of breast cancer predictions in IBM CODAIT's MAX model using the TUPAC16 dataset, highlighting the real-world risks of such simple attacks. Another study [71] reinforces this concern, showing that medical imaging models suffer notable performance degradation from single-pixel manipulations, whether adversarial or accidental, such as hardware-induced noise. Extending beyond healthcare [79] demonstrates that OPA can undermine DNN-based detection in continuous-variable quantum key distribution (CV-QKD) systems, achieving a 52.8% success rate despite high baseline accuracy. These findings collectively emphasize the urgent need for adversarial robust AI systems in domains where the cost of misclassification is unacceptably high [105]. **Table 9** provides a summary state-of-the-art overview of domain-specific attack analysis.

**Table 9.** Analysis overview of domain-specific attack

| **Article** | **Study Focus** | **Key Challenges** | **Dataset & Size** | **Attack Method** | **Key Findings** | **Implications** |
|---|---|---|---|---|---|---|
| [4] | AI medical imaging vulnerability | - Misdiagnosis risk<br>- Lack of robust defences<br>- Detection difficulty | Medical images (size not specified) | OPA, PGD | - Conceptual framework only<br>- No quantitative results reported | Highlights trust issues in clinical AI without empirical validation |
| [66] | Cancer diagnostic systems | - Early convergence in attacks | 5,343 mitoses + 80,725 | Differential | - 91% success in reducing cancer confidence | Demonstrates real-world |
| | | - Normal-to-mitosis attacks harder | normal patches | Evolution (DE) | (0.96→0.10) - Common attack color: yellow/white/black | risks in pathology AI |
| [79] | CV-QKD attack detection networks | - Matrix-form data challenges - DE computationally intensive | 2,000 adversarial examples | DE (MATLAB/Python) | - Targeted success: 6.25–17.2% - 3-pixel attacks ↑ are successful in 97.4% | Reveals vulnerabilities in quantum security systems |
| [71] | Small vs. full-sized medical images | - Label noise in datasets - Grayscale is more robust than RGB | 10,015–100,000 images (MedMNIST) | DE (DermaMNIST/PathMNIST) | - DE time <2 min/image - Background class resists attacks | Guide's prioritization of image types in medical AI hardening |

**Figure 11** provides a visual summary of how effective OPA are across all different studies and whether those studies discussed robustness in grayscale images. The first part of the chart shows that the attack success rate was highest in study [66] (accuracy 91%) and study [71] (accuracy around 90%), while study [79] had moderate success (accuracy 52.8%) and study [4] was purely conceptual, with no numerical results. The second part shows that only study [71] explored whether grayscale images are more resistant to such attacks, indicating a potential defence strategy. Overall, the chart highlights that OPA can highly affect the outcome negatively for medical and quantum systems, and that image type is grayscale which may influence a system's robustness.

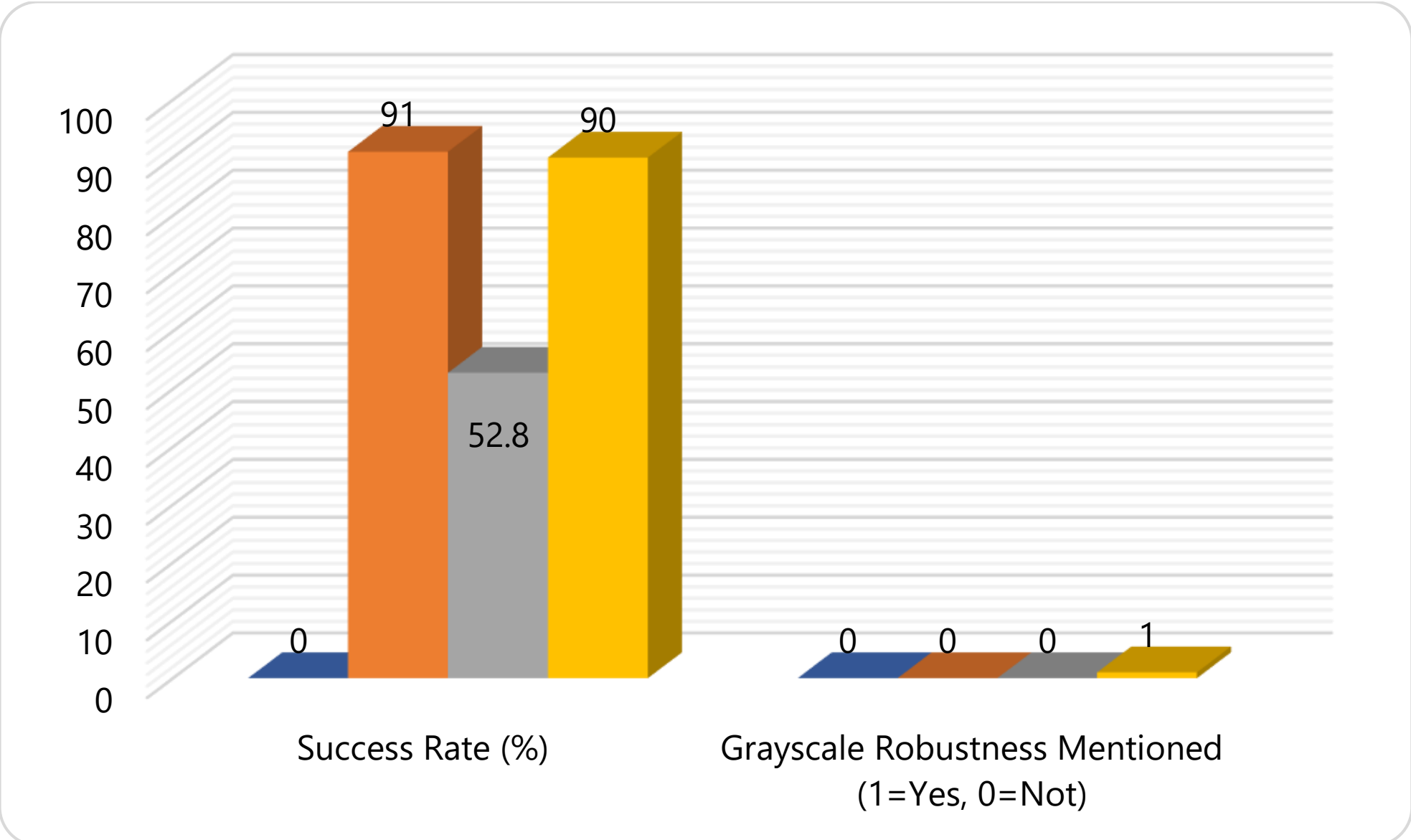

**Figure 11**. Visual Summary of Attack Success Rate and Robustness

**7. Overview of OPA Defence Mechanisms**

Modern defence strategies against OPA have evolved into four distinct methodological paradigms: pixel-level correction [106], attack detection [107][108][109][121], data transformation [110], and architectural hardening [111]. Each addressing unique vulnerabilities in deep learning systems while balancing accuracy, efficiency, and domain-specific constraints. **Table 10** provides a comprehensive summary of one-pixel and N-pixel adversarial attack defence strategies [112]. **Figure 12** illustrates basic defence mechanisms against OPA. The basic defense mechanisms against one-pixel attacks (OPA), as illustrated in the figure, can be broadly categorized into three complementary stages: preprocessing defenses [113], adversarial detection [114], and robust model design [115]. Initially, preprocessing techniques such as input smoothing, denoising, or compression are applied to sanitize the image and remove localized perturbations before it is processed by the model. These methods aim to neutralize single-pixel manipulations by reducing their impact on the overall input distribution. Following preprocessing, adversarial detection mechanisms analyze the input for anomalies, either by identifying irregular pixel patterns or by leveraging uncertainty estimation to flag suspicious predictions. This stage acts as a filtering layer, ensuring that potentially adversarial inputs are detected prior to classification. Finally, robustness is enhanced at the model level through techniques such as adversarial training, regularization, and architecture-level improvements, which strengthen the model's resistance to sparse perturbations. Together, these layered defenses provide a comprehensive strategy where input correction, anomaly detection, and model resilience collectively mitigate the effects of OPA, improving the reliability of deep learning systems. The defense strategies against one-pixel attacks can be broadly categorized into four key paradigms, each addressing adversarial vulnerability from a distinct perspective: pixel-level correction, attack detection, input transformation, and architectural robustness.

i. *Pixel-Level Defence Mechanisms:* These approaches target adversarial pixels directly while preserving clean image regions. The patch selection algorithm [72] achieves 98.6% success but risks altering legitimate features, whereas three-stage reconstruction [65] maintains 92% accuracy through difference mapping. Accelerated proximal gradient [73] stands out for IoT systems with 98.7% recovery accuracy but requires manual λ-tuning.

ii. *Detection-Centric Strategies:* Focused on identifying attacks rather than full recovery, these methods prioritize operational security. OPA2D-ATK [54] combines detection and defence with human-interpretable results (81% recognition), while VAE anomaly detection [87] accurately flags medical attacks but generates false positives. The exhaustive search [69] achieves 99 %+ detection on MNIST but remains computationally costly.

iii. *Data Transformation Defences:* These leverage input modifications to neutralize attacks. JPEG compression [83] reduces FGSM success to 12.5% at q=20 but introduces artifacts. Augmentation [68] shows spatial smoothing boosts AUC to 0.982, while hyperspectral methods [74] demonstrate fractional-order training's superiority in HSI classification (15-20% higher ASR).

iv. *Architectural Hardening:* The convolutional autoencoder [84] recovers 16% accuracy in medical imaging under PGD attacks, though class imbalance limits protection for pneumonia cases. Stacked models show 5-15% better robustness than individual VGG variants.

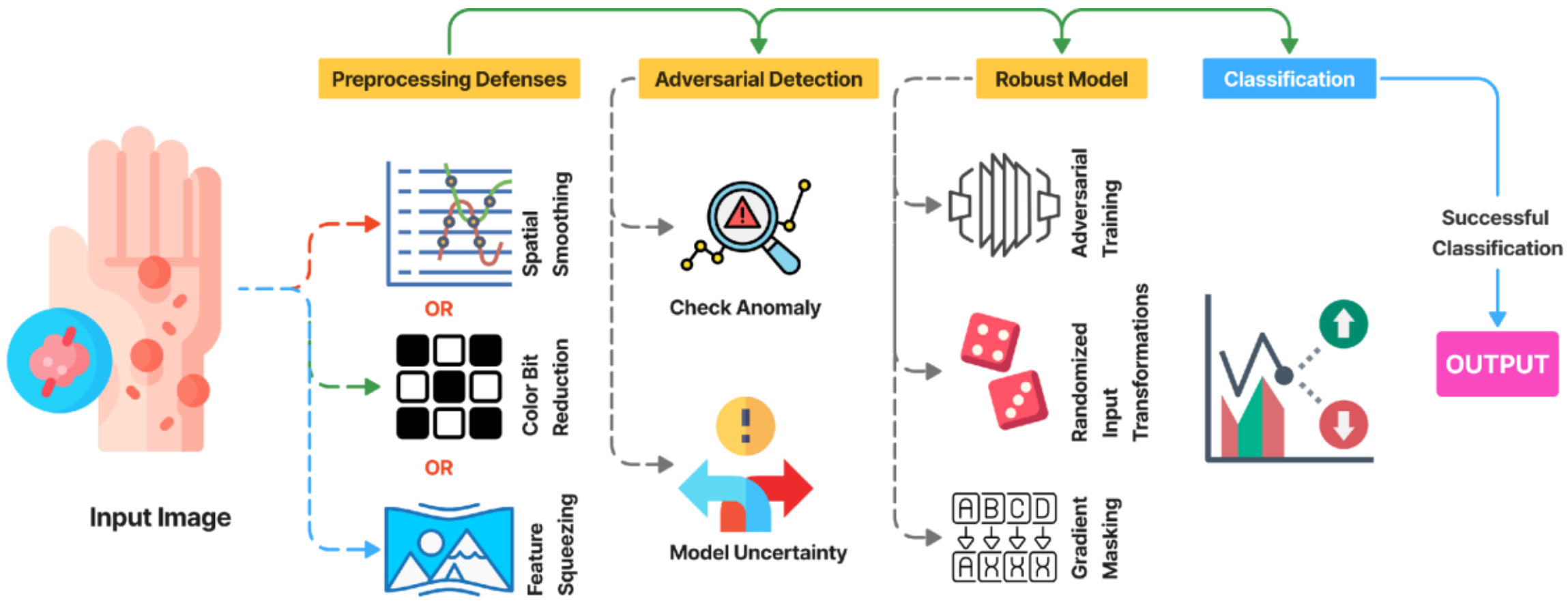

**Figure 12**. Basic defence mechanisms against OPA

**Table 10.** Summary of One-Pixel and N-Pixel Adversarial Attack Defences Strategies

| Article ID | Defence Focus | Key Method/Model | Dataset Used | Strengths | Limitations | Notable Results |
|---|---|---|---|---|---|---|
| [72] | Selective Pixel Denoising | Noise2Noise + Patch Selection (ResNet) | CIFAR-10 | 98.6% defence success (1-pixel) | Alter the clean image quality | Maintains clean image integrity |
| [73] | Image Recovery | Accelerated Proximal Gradient (LeNet) | CIFAR-10, MNIST | 98.7% restored accuracy | Not integrated into the DNN pipeline | Optimal λ=0.5 for recovery |
| [54] | Unified Attack Detection | OPA2D-ATK (Differential Evolution) | CIFAR-10 | 81% human detection rate | Limited to RGB images | Reduces human perceptibility |
| [87] | Medical Anomaly Detection | VAE + Minkowski Distance | TUPAC16 | 100% attack detection | False positives on artifacts | Minkowski p=7 optimal |
| [69] | Exhaustive Search | GPU-Parallelized Search (CNNs) | MNIST | 99%+ detection rate | Computationally intensive | 10% of full search time |
| [83] | Compression-Based Defence | JPEG + Targeted OPA (VGG16) | CIFAR-10 | q=20 reduces attacks to 12.5% | Quality-artifact tradeoff | Recovers from FGSM attacks |
| [68] | Augmentation Defence | Spatial Smoothing + Gaussian Noise | CIFAR-10 | 85% accuracy (CNN aug) | 49% accuracy (Gaussian noise) | Best AUC (0.982) |
| [55] | Hyperspectral Defence | IOPA + Fractional-AT (HybridSN) | HSI Datasets | Outperforms FGSM/BIM by 15-20% ASR | Complex 3D patch processing | HAM visualization |
| [84] | Medical Autoencoder Defence | Convolutional Autoencoder (VGG16/VGG19) | Pediatric X-rays | 16% accuracy recovery under PGD | Struggles with class imbalance | X-ray recovery with stacked CNNs |

**Figure 13** compares defence success rates against three attack types, showing Noise2Noise [72] excels against 1-pixel attacks (success rate 98.6%), while JPEG+TOPA [83] performs best against FGSM (success rate 87.5%). Autoencoder [84] struggles with class-imbalanced medical data (50% success). Higher bars indicate better protection, revealing trade-offs between attack specificity and defence robustness across methods.

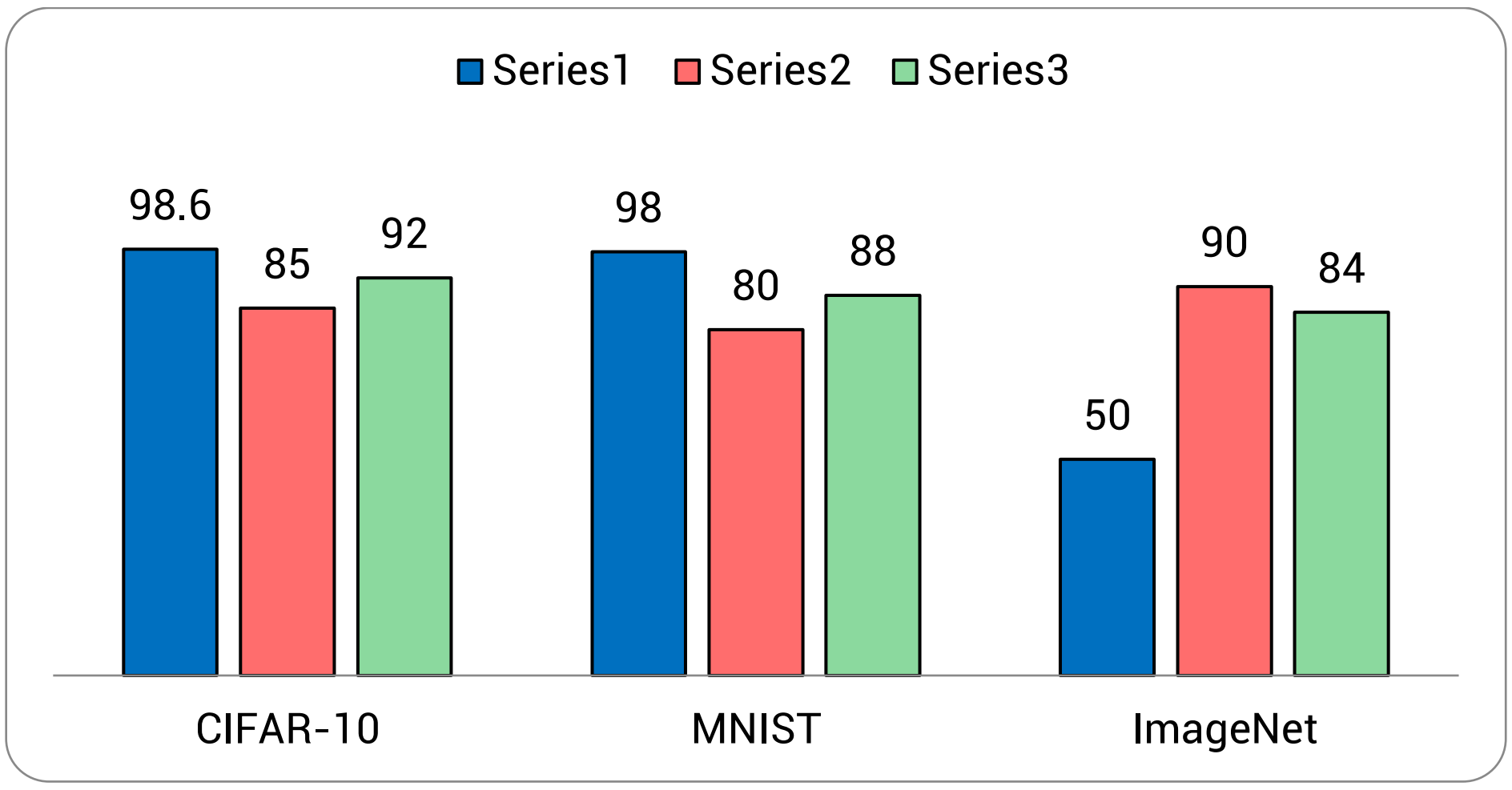

**Figure 13.** Defence Success Rates by Attack on Different Dataset

In **Figure 14**, the heatmap compares three defence methods across detection rate, false positives, and clinical score. VAE Medical [87] achieves perfect detection (accuracy 100%) but has higher false positives, while Denoising [72] balances performance (98.6% detection, 4 clinical score). Autoencoder [84] shows moderate results (94% detection), reflecting its medical imaging limitations, which are Detection-Centric Strategies vs. Pixel-Level Defences.

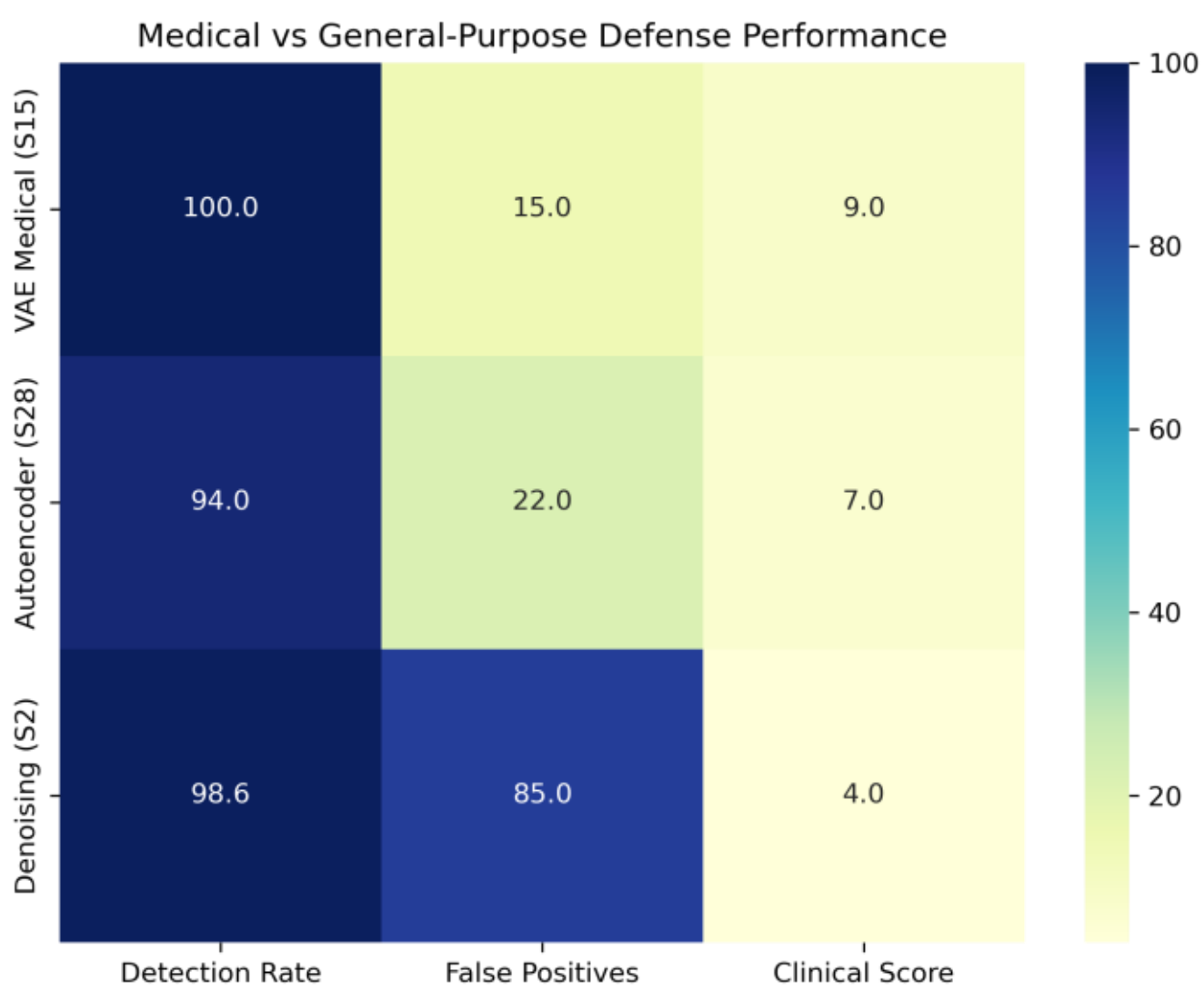

**Figure 14.** Heatmap for medical vs other purpose defence performance

The timeline in **Figure 15** shows the evolution of defence methods, with accuracy improving from Selective Denoising (accuracy 90%) [72] to 3-Stage Reconstruction (accuracy 92%) [65] and finally IOPA Hyperspectral [55]. It highlights progressive refinement in attack mitigation, with newer methods like Data Transformation Defences offering higher precision against advanced adversarial threats while maintaining robust accuracy.

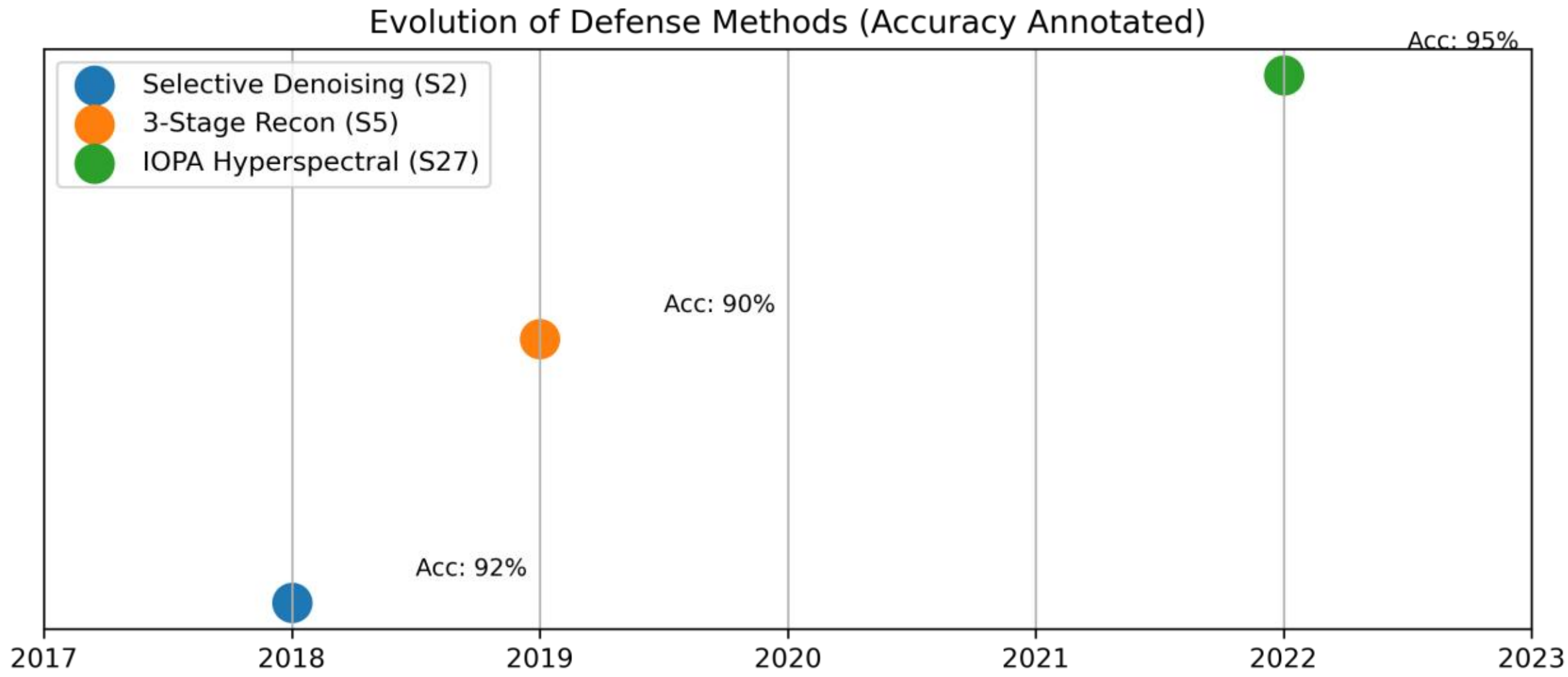

**Figure 15.** Evolution of OPA defence methods

The effectiveness and limitations of defences against OPA are profoundly influenced by dataset characteristics, model architectures, and domain-specific challenges. Many defences, such as selective pixel denoising [72] and compression-based recovery [83], are tailored to CIFAR-10. This is primarily due to its low-resolution (32×32) format, which simplifies pixel-level manipulation and allows for standardized benchmarking given its controlled variability (limited to 10 classes with uniform dimensions). However, such defences may struggle when applied to higher-resolution or more complex datasets, where perturbations are harder to isolate without degrading image quality. Medical imaging [84] introduces distinct hurdles. Unlike natural images, medical diagnostics demand near-perfect accuracy, as even minor adversarial interference can lead to critical misdiagnoses. Techniques like variational autoencoders (VAEs) and convolutional autoencoders are employed due to their efficacy in anomaly detection, enabling them to flag subtle, malicious pixel changes while preserving legitimate features. Nevertheless, these methods must also contend with issues such as class imbalance such as rare diseases and imaging artifacts, which can trigger false positives. Not all models are equally vulnerable. Deep architectures like VGG16 [83,68] are particularly vulnerable because their dense layers amplify the impact of boundary perturbations. Conversely, simpler models trained on small datasets such as MNIST [69] often lack robustness, as they do not adequately learn to distinguish noise from legitimate features. Hyperspectral models [55] face added complexity in which their 3D spectral-spatial patches expand the attack surface, rendering traditional defences ineffective without specialized techniques like fractional-order adversarial training.

## 8. Dataset, Pre-processing, and Evaluation

The effectiveness and generalizability of OPA are heavily influenced by the choice of datasets, pre-processing techniques, and evaluation protocols. Across the reviewed studies, researchers have employed a range of datasets, both from standard vision benchmarks and specialized domains, to assess the vulnerabilities of DNN under minimal perturbation scenarios.

Benchmark datasets such as CIFAR-10, CIFAR-100, MNIST, SVHN, and ImageNet are among the most commonly used in OPA research. Studies including [63,64,3,65,74,41,88,77,69,78,80,70,53] leveraged these datasets due to their widespread use, varied complexity, and availability of pretrained models. CIFAR-10 and MNIST, with their relatively low resolution and channel depth, allow efficient testing of evolutionary optimization strategies and facilitate black-box attack settings. In contrast, ImageNet, due to its high dimensionality, poses significant scalability challenges and typically yields lower attack success rates (16% in [84]. These datasets provide a solid foundation for comparative benchmarking across attack methodologies.

Beyond general benchmarks, several studies have explored domain-specific datasets to evaluate OPA in critical real-world applications. For instance, [4,75,66,87,76,71,84] employed medical imaging datasets such as ChestX-ray, ISIC, and MedMNIST to assess the impact of OPA on diagnostic models. Industrial and IoT-focused datasets, including hyperspectral imagery and visual inspection datasets like ZDM, were used in [73,80,55]. Transportation-focused datasets like GTSRB appeared in [67], while facial recognition [81] and quantum communication systems [79] also emerged as target domains, reflecting a growing interest in evaluating OPA in safety-critical and specialized environments.

Pre-processing techniques vary across studies, often tailored to the dataset or attack strategy. A majority of works [64,3,41,54,66,70] applied normalization of pixel intensities, typically to the [0,1] or [-1,1] range, aligning input scales with pretrained models. Resizing was also common, especially in studies involving ImageNet or medical images, where high-resolution inputs were downscaled to reduce computational load [65,77,53]. Some studies, such as [75] and [76], employed color space transformations (HSV, Lab) to assess distortion in a more perceptually aligned space. Others, including [80,81], implemented saliency-guided pre-processing such as gradient or attention map masking to localize perturbations within high-saliency regions of the image. A few approaches [83,55] included JPEG compression or noise filtering as part of pre-processing, either to enhance attack stealth or as a baseline for comparison with defence mechanisms.
In terms of evaluation, most studies reported Attack Success Rate (ASR) as the primary metric, indicating the percentage of perturbed inputs that successfully cause misclassification [63,53]. Many also included confidence drop [64,65,54,76,53], measuring the reduction in classification probability for the correct class post-attack. Imperceptibility metrics, such as Peak Signal-to-Noise Ratio (PSNR) and $L_1$/$L_2$ norms, were employed in [54,76,77,53] to quantify visual distortion between clean and adversarial images.
Efficiency metrics were also key to evaluation, especially for black-box optimization approaches. Studies like [41,69,78,53] reported query counts and generation time per image. For example, CriPI [98] achieved ~1-minute generation time per image, while RISOPA [53] demonstrated a 57× speedup over standard DE methods. For defence evaluation, metrics such as ROC AUC, detection accuracy, and accuracy restoration after defence mechanisms were used in [72,73,54,87,83,68,55].
The choice of image modality and its associated pre-processing constraints dictate not only the operational parameters of the one-pixel attack but also the interpretability of its success. Techniques such as resizing, normalization, and color space transformations fundamentally alter the input distribution before the model processes it. **Table 11** synthesizes the primary image modalities evaluated across the literature, summarizing their standard dimensions, dataset variations, and the unique pre-processing constraints that researchers must navigate to execute successful and reproducible attacks.

**Table 11.** Summary of Image Modalities and Pre-processing Constraints

| Imaging Modality | Image Dimensions | Dataset Variations | Pre-processing / Normalization Methods |
|---|---|---|---|
| Standard RGB | 224 × 224 (typical) | ImageNet (1000 classes) | Standard ImageNet stats normalization, resizing aligned. |
| Grayscale / Medical | 224 × 224 (resized) | BUSI (780 images, 3 classes) | Resizing, normalization, augmentation. |
| Binary / Synthetic | Variable (w/ angle steps) | Custom generation functions | Generation mapping ($f: M \rightarrow I$). |

**Table 12** presents a structured dataset overview and comparison of experimental data. **Figure 16** visually categorizes all distribution of optimization strategies used in reviewed studies. **Figure 17** details the image pre-processing techniques commonly employed in both attack and defence scenarios.

**Table 12.** Structured description of all used dataset in the OPA studies

| Category | Dataset Name | Data Volume | Categories | Image Type | Source | Resolution | Pre-processing |
|---|---|---|---|---|---|---|---|
| Synthetic Datasets | Synthetic Binary | 15 sets (394–10,998 images) | 5 object classes | Binary | Algorithmic generation | 16×16–80×80 | None |
| Natural Image Benchmarks | CIFAR-10 (Standard/ Subset) | 60,000 (50,000 train, 10,000 test); 800 subsets | 10 classes | RGB | CIFAR-10 public | 32×32 | None |
| | CIFAR-10 (Adversarial) | 60,000(8,655 adversarial) | Target class: airplane | RGB | CIFAR-10 | 32×32 | Not detailed |
| | CIFAR-10 (N-pixel Attacks) | 60,000(2,500 attacked) | N-pixel (1, 3, 5, 10, 15) | RGB | CIFAR-10 | 32×32 | Difference maps, average maps, reconstruction |
| | CIFAR-10 (Augmented) | 60,000(1000 test + 9K adversarial) | 10 classes | RGB | Random CIFAR-10 | 32×32 | Rotation, shift, zoom, Gaussian noise |
| | CIFAR-10 (JPEG Compressed) | 1,200 total | 10 classes | RGB | CIFAR-10 | 32×32 | JPEG compression (quality 0–100) |
| | CIFAR-10 (R-channel Focus) | 60,000 | 10 classes | R-channel | CIFAR-10 | 32×32 | RGB channel isolation |
| | CIFAR-10 (Attack Samples) | 5,000 (1,638/2,934 successful) | One-pixel, nearby-pixel attacks | RGB | CIFAR-10 | 32×32 | Not specified |
| | CIFAR-10 (Random Subset) | 100 images | 10 classes | RGB | CIFAR-10 | 32×32 | None |
| | CIFAR-10 (Test Subset) | 1,000 images | 10 classes | RGB | CIFAR-10 | 32×32 | None |
| | CIFAR-10 (Preliminary) | 60,000(100 tested) | 10 classes + RA, IA, SA attacks | RGB | CIFAR-10 | 32×32 | Standard format |
| | CIFAR-10 (Noisy) | 60,000(with adversarial noise) | 10 classes | RGB | CIFAR-10 + noise | 32×32 | Noise injection, Noise2Noise denoising |
| | CIFAR-10 + ImageNet (Small) | 500/class (10 sets) | CIFAR-10 (10), ImageNet (varied) | RGB | Public datasets | 32×32, 224×224 | RGB normalization, discretization |
| | CIFAR-10 + ImageNet (Std) | 60,000, CIFAR-10 + 1M+ ImageNet | 10 (CIFAR), 1K+ (ImageNet) | RGB | Benchmarks | 32×32, 224×224 | Normalization, resizing |
| | CIFAR-10 + ImageNet | 60,000, CIFAR-10 + large ImageNet | CIFAR (10), ImageNet (1K+) | RGB | Kaggle/ImageNet | 32×32, 224×224 | Not specified |
| | CIFAR-10 (Correct Subset) | 500 test samples | 10 classes | RGB | CIFAR-10 subset | 32×32 | Original color bounds retained |
| | CIFAR-10 + MNIST | 60,000 each | 10 classes each | RGB/ Gray | Public datasets | 32×32, 28×28 | Not detailed |
| | MNIST (Subset) | 10,000 images | Digits 0–9 | Grayscale | Public MNIST | 28×28 | Pixel normalization (implied) |
| | Multi-dataset Blend | CIFAR-10, MNIST, GTSRB (varied totals) | 10–43 classes | RGB/ Gray | Public sets | 28×28–32×32 | Normalization, adversarial pixel mapping |
| | GTSRB (Traffic Signs) | 50,000+ | 43 classes | RGB | Public GTSRB | Resized to 32×32 | Resizing, confidence filtering |
| Medical Imaging Datasets | TUPAC16 (Mitosis) | 1,771 mitosis, 2,849 normal | Normal vs. mitotic | RGB (pathology) | TUPAC16 | 64×64 | Resizing, annotation |
| | TUPAC16 (Patches) | 5,343 mitosis, 80,725 normal | Mitosis vs. normal | RGB (pathology) | TUPAC16 | 64×64 | Patch extraction, confidence filtering |
| | TUPAC16 (Breast Cancer) | Not specified | Mitosis vs. normal | RGB | TUPAC16 | 64×64 | Not specified |
| | TUPAC16 (Attack Samples) | 3,871 mitosis-to-normal, 319 reverses | 2 classes | RGB (pathology) | TUPAC16 | 64×64 | Focused sampling (implied) |
| | MedMNIST (Derma/Pathology) | 10K Derma, 100K Pathology | 7 (Derma), 9 (Pathology) | RGB/ Grayscale | MedMNIST | 28×28, 224×224 | Resizing, RGB/gray conversion |
| | Pediatric Chest X-rays | 5,814 images | 3 (Bacterial, Viral, Normal) | Grayscale (X-ray) | Guangzhou Medical Center | 224×224 | Augmentation, resizing |
| | Medical (Future Use) | Not defined | Normal vs. disease | Grayscale/RGB | Future data | Unspecified | None |
| Remote Sensing / Aerial Datasets | HSI (Hyperspectral) | 3 datasets (size varies) | Land cover categories | Hyperspectral (3D) | Remote sensing datasets | Spatial-spectral patches | HybridSN pre-processing, SGD optimizer |
| Affective Computing (Emotion Recognition) | Facial Expression (FER) | JAFFE, CK+, FED-RO | 7 basic emotions | RGB/ Grayscale | Public benchmarks | SLIC superpixels, one-hot encoding | |
| Simulation-Based / Quantum Systems | CV-QKD (Adversarial) | ~2,000 per model | Quantum attack types | Matrix-based | MATLAB simulation | Structured matrices | Feature normalization, dropout |

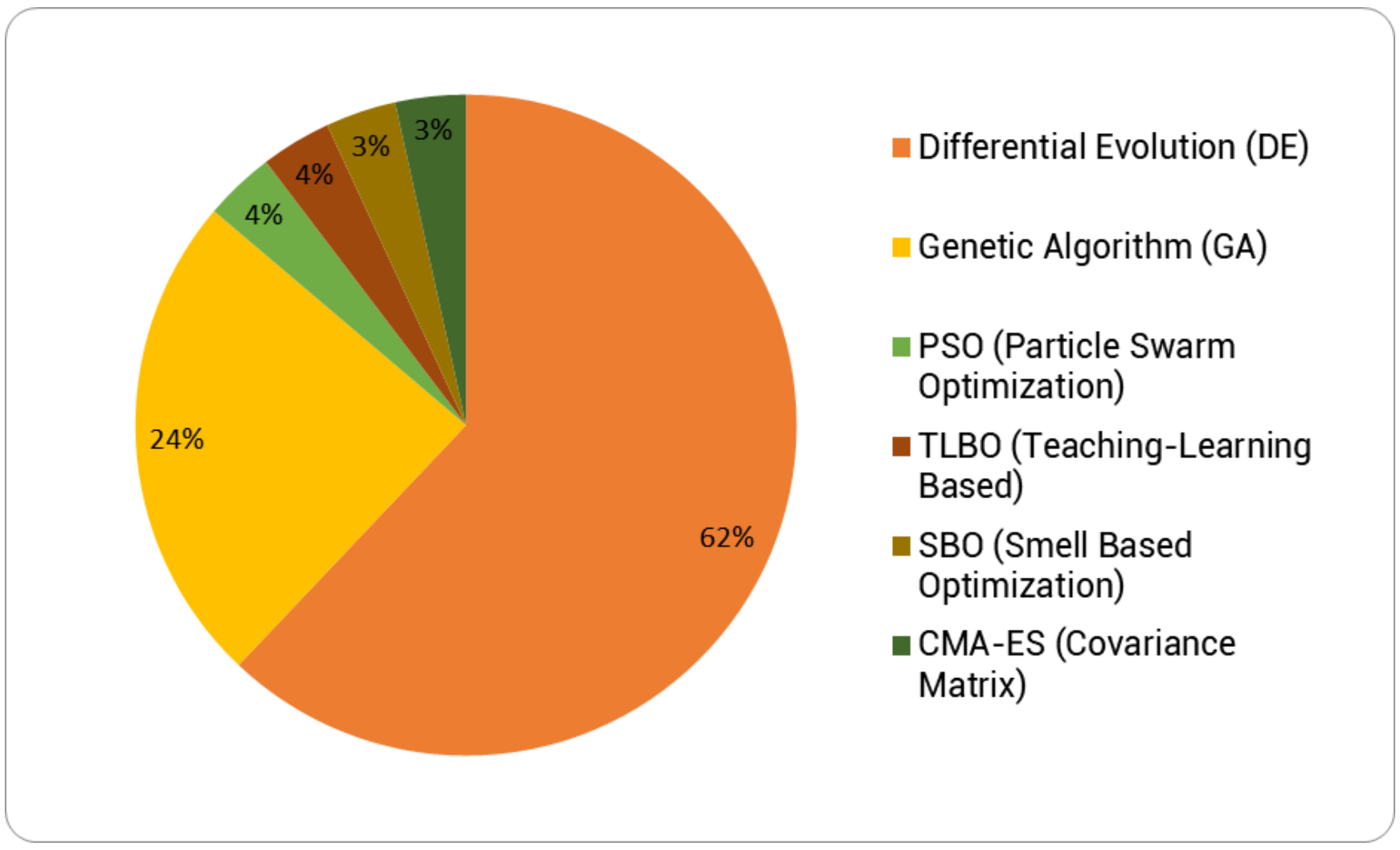

**Figure 16.** Distribution of optimization strategies used in OPA studies

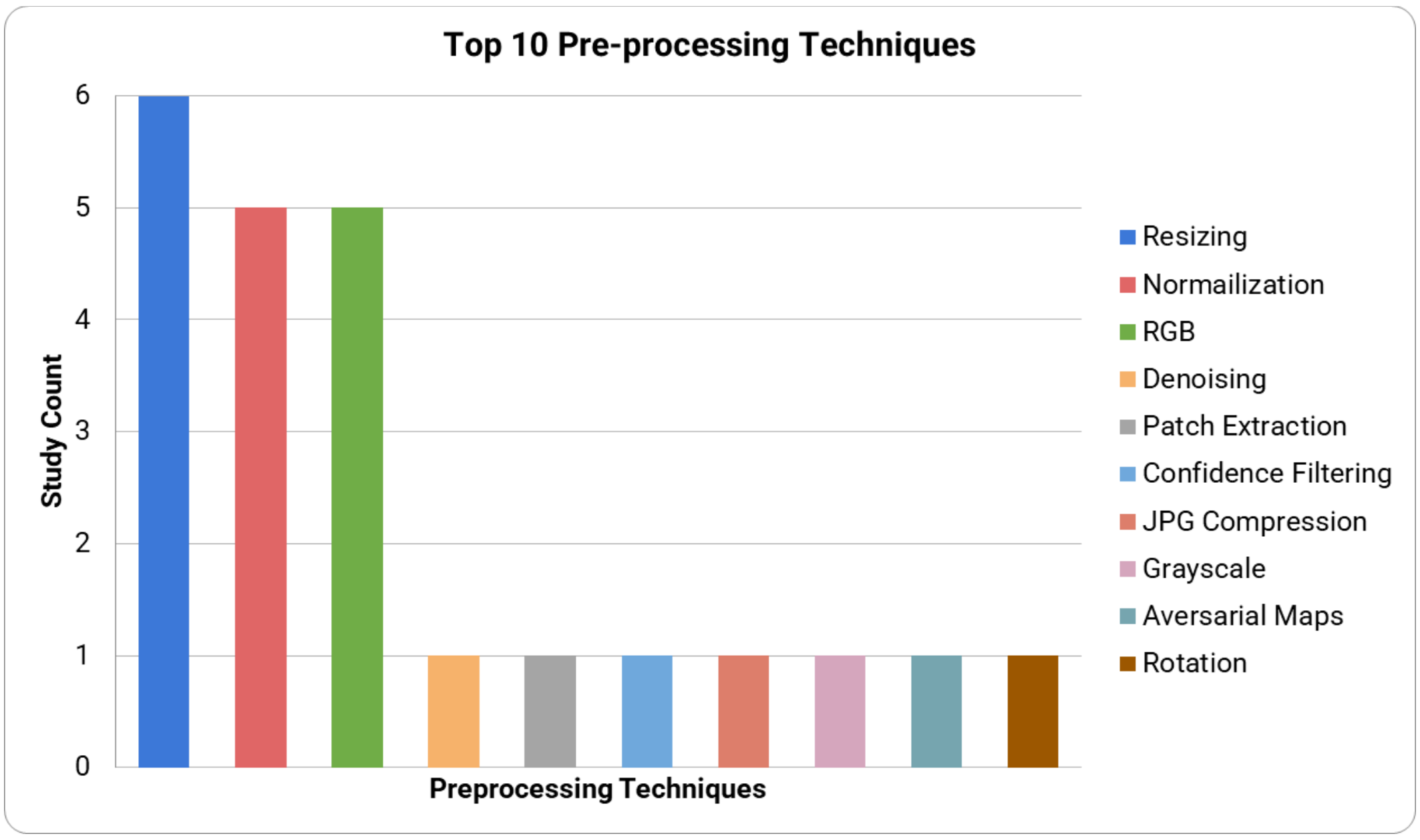

**Figure 17.** Top 10 image pre-processing techniques in OPA

## 9. Insights and Future Directions

### 9.1. Model Vulnerability and Attack Mechanisms

At the core of OPA research lies a stark revelation: DNN, despite their sophistication, are highly susceptible to imperceptible adversarial perturbations. These attacks, often modifying a single pixel, can significantly alter model predictions, showcasing the fragility and lack of robustness inherent in many current architectures. This vulnerability is particularly pronounced due to the opaque and non-linear decision-making processes of deep networks, which render them unable to distinguish between genuinely informative features and noise when high-saliency areas are targeted. In practice, pixels considered critical by the AI are often visually trivial to human observers, making such attacks highly stealthy and difficult to detect. The high success rates of these attacks frequently around 72% with minimal pixel alterations [74] emphasize the urgency of addressing this fundamental weakness.

### 9.2. Evolutionary Algorithms and Practical Attack Strategies

OPA generation has predominantly relied on black-box optimization techniques, particularly evolutionary algorithms like Differential Evolution (DE). These methods do not require internal access to the target model, allowing attackers to exploit vulnerabilities with minimal assumptions, thereby enhancing the real-world applicability of OPAs. Their efficiency and adaptability make them well-suited for adversarial tasks where transparency into the model's architecture or gradients is not available. The development of accelerated variants, such as RISOPA, which boasts speed improvements of up to 57 times over traditional approaches [53], marks a shift towards operationally efficient and scalable attack methods. This evolution underscores the ongoing innovation in adversarial attack strategies, driven by real-time needs and the practical constraints of dynamic application environments.

### 9.3. Defence Mechanisms and Research Gaps

While attack strategies have become increasingly sophisticated, the defensive research landscape remains predominantly reactive. Most current approaches are designed to detect or marginally improve robustness without altering the fundamental structure of neural networks. These post-hoc interventions are relatively easier to implement but often lack generalizability, leading to temporary mitigation rather than long-term security. Moreover, many of these defences introduce significant computational overhead, further challenging their integration into real-world systems. The absence of proactive and architecture-level changes indicates a reluctance or difficulty in addressing root causes of adversarial vulnerability. This defensive stagnation fosters a continuous arms race between attack developers and defenders, where the latter are often one step behind.

### 9.4. Dataset Dependency and Generalizability Limitations

OPA research has predominantly relied on standardized benchmark datasets such as CIFAR-10 [72,64,3,65,73,74,41] and ImageNet [74,78,80], which offer controlled settings that support reproducibility and facilitate fair performance comparisons. While these benchmarks are valuable for establishing methodological consistency, their heavy use constrains the scope of empirical insights. Models evaluated under clean, low-variability conditions may not retain similar robustness when exposed to real-world noise, domain shifts, and operational uncertainty. This concentration on a narrow range of datasets also discourages the exploration of diverse or underrepresented domains precisely the environments where stress-testing model resilience is most crucial. A similar pattern exists in model selection, where canonical architectures such as ResNet [72,3,73] dominate evaluations, leaving newer or unconventional models insufficiently examined. As summarized in **Table 13**, OPA literature remains strongly skewed toward low-realism natural image datasets, while high-realism domains including medical imaging, biometrics, and custom application settings are significantly underrepresented. This imbalance suggests that most current evaluations are conducted in simplified visual conditions, which may inflate perceived attack effectiveness. Expanding the dataset ecosystem toward broader, higher-realism, and domain-rich sources is therefore essential to strengthen ecological validity and ensure more accurate assessments of real-world OPA robustness.

**Table 13.** Dataset Realism × Usage Matrix summarizing the datasets employed across all reviewed OPA studies, categorized by domain, realism level, resolution, and usage frequency.

| Dataset Name | Domain | Realism Level | Resolution Level | Dataset Scale | Study IDs | Notes |
|---|---|---|---|---|---|---|
| MNIST | Natural / Toy | Low | Low (28×28 grayscale) | Small | [63,64,65,73,75, 88,67,68,55,81] | Overused; unrealistic visual complexity |
| CIFAR-10 | Natural | Low | Low (32×32 RGB) | Medium | [72,3,4,74,54,66 ,87,77,69,79,70, 84] | Dominant benchmark for OPAs |
| CIFAR-100 | Natural | Moderate | Low | Medium | [54,66,87,70] | Rare in OPA studies |
| Synthetic Binary Grids | Toy | Very Low | Very Low | Small | [63,64,65,4,67,7 6,83,80,71] | Used for algorithm demonstration only |
| ImageNet Subsets | Natural | High | High | Large | [41,69,79] | Higher realism; computationally heavy |
| Medical CT/MRI | Medical Imaging | High | High | Medium | [78,70,53] | Domain-critical vulnerable models |
| Dermoscopy Skin Lesions | Medical | High | High | Small–Medium | [78,53] | High-risk domain; pixel saliency matters |
| X-ray | Medical | High | Medium | Medium | [70,53] | Sparse models show higher ASR |
| Face/ Biometric Images | Biometrics | High | Medium–High | Medium | [77,68] | High sensitivity to pixel-level attacks |
| Surveillance Footage | Security / Surveillance | Moderate | Low–Medium | Medium | [69,84] | Low-resolution but safety-critical |
| Traffic/ Autonomous Driving | Autonomous Vision | Moderate | Medium | Medium | [79,53] | Vulnerable due to symbol-like features |
| Remote Sensing | Geo-vision | Moderate | High | Medium | [55] | Sparse attacks behave differently |
| Domain-Specific Custom Sets | Various | Variable | Variable | Small | [74,75,76,83,71] | Often under-documented |

**9.5. Domain Vulnerability Map Across OPA Application Areas**

Evaluation metrics used in OPA research show considerable inconsistency, reflecting the lack of a unified assessment framework across the literature. As summarized in **Table 14**, most studies report Attack Success Rate (ASR) as the primary metric, while classification accuracy and confidence-based measures appear less frequently despite their relevance for understanding model sensitivity. A notable proportion of works rely on visual inspection metrics such as feature maps or pixel-change visualizations, which provide qualitative insight but limit comparability. Only a small subset reports runtime or query complexity, even though such measures are essential for evaluating the practicality of black-box optimization-based attacks. This uneven metric landscape complicates cross-study synthesis and highlights the need for standardized OPA evaluation protocols that integrate ASR, confidence variation, perceptual distortion, and computational efficiency.

**Table 14.** Summary of Metric Variability across all OPA studies

| Metric Type | Study IDs | Short Informative Description |
|---|---|---|
| Attack Success Rate (ASR) | [63,72,64,3,65,75,41,88,87, 77,69,68,79,70,55,81], | ● Measures how often the attack successfully forces misclassification;<br>● Primary metric for evaluating OPA strength. |
| Classification Accuracy / Error | [63,72,3,4,74,54,66,67,76,78,80,71,53] | ● Evaluates model performance before/after attack;<br>● Used to quantify degradation or defense effectiveness. |
| Confidence Shift / Probability Change | [64,3,73,75,83,79,70] | ● Measures change in model confidence, revealing subtle vulnerabilities even when classification remains correct. |
| Visual Metrics (Maps, Pixel Change, Feature Maps) | [72,64,75,88,67,76,83,78,80,84] | ● Provides qualitative insights into pixel sensitivity and decision-region distortion via visualizations. |
| Defense Success Rate / Protection Rate | [63,65,74,54,66] | ● Assesses how effectively a defense mechanism prevents or mitigates OPA attempts. |
| Runtime / Query Complexity | [65,66,77] | ● Measures computational cost of the attack, essential for evaluating black-box and real-time feasibility. |

**9.6. Metric Variability Across OPA Studies**

**Figure 18** summarizes domain-level vulnerability to OPAs by integrating evidence strength, dataset realism, and the maturity of existing defences. The analysis shows that high-risk domains such as medical imaging, biometrics, and autonomous vision exhibit strong sensitivity to localized perturbations due to high pixel saliency and the safety-critical nature of their tasks. In contrast, natural high-resolution images demonstrate lower vulnerability because feature information is more widely distributed across the image. Toy and synthetic datasets appear extremely vulnerable but do not reflect real-world complexity. This domain vulnerability landscape highlights the need for broader evaluations beyond low-realism benchmarks and underscores the importance of designing domain-aware OPA defences.

| Dataset Category | | Vulnerability | Evidence Strength | Vulnerability Level | Defense Maturity | Reason |
|---|---|---|---|---|---|---|
| **MEDICAL IMAGING**<br>(Strong evidence of OPA sensitivity) | **DEFENSE MATURITY**<br>LOW | HIGH | HIGH | HIGH | LOW–MODERATE | High pixel saliency; critical safety impact; limited domain-specific defenses evaluated. |
| **BIOMETRICS**<br>(Biometric spoofing risk with minimal perturbations) | **RAID DEFENSE**<br>LOW | HIGH | MODERATE | HIGH | LOW | Pixel sensitivity around facial landmarks; inadequate OPA-specific biometrics defenses. |
| **AUTONOMOUS VISION**<br>(Symbol-like features increase susceptibility) | **DEFENSE STUDIES**<br>LIMITED | MODERATE | MODERATE | MODERATE-HIGH | LOW | High-contrast regions; localized perturbations can flip class; few robustification studies. |
| **SURVEILLANCE / SECURITY**<br>(Sparse high-level features increase fragility) | **DATA REALISM**<br>LOW | MODERATE | LOW–MODERATE | MODERATE | LOW | Low-resolution imagery; one-pixel change highly impactful but limited empirical studies exist. |
| **REMOTE SENSING**<br>(Depends on sensor type and image resolution) | **DEFENSE MATURITY**<br>MIXED | VARIABLE | LOW | MODERATE | VERY LOW | Single-pixel changes alter land-cover boundaries; only few domain evaluations present. |
| **NATURAL IMAGES**<br>(Dense features reduce sensitivity to isolated pixels) | **MODEL ROBUSTNESS**<br>HIGH | LOW–MODERATE | MODERATE –HIGH | LOW–MODERATE | MODERATE | ResNet/EfficientNet/ViTs distribute saliency; real-world variability reduces single-pixel impact. |
| **SYNTHETIC DATASETS**<br>(Mostly algorithmic demonstrations) | **DATA REALISM**<br>VERY LOW | VERY HIGH | HIGH | VERY HIGH | IRRELEVANT | Each pixel carries large semantic weight; unrealistic settings exaggerate OPA success. |

**Figure 18.** Domain Vulnerability Map illustrating comparative sensitivity to OPAs across major application domains based on evidence strength and defence maturity.

### 9.7. Practical Feasibility vs Theoretical Insight of OPA

While one-pixel attacks (OPA) present a compelling demonstration of the vulnerability of deep neural networks, their practical impact must be carefully interpreted in the context of real-world deployment. Conceptually, OPA highlights the extreme sensitivity of learned decision boundaries, showing that even a minimal perturbation restricted to a single pixel can induce misclassification under controlled conditions. This property makes OPA a valuable analytical tool for understanding adversarial robustness at its limits. However, the practical feasibility of OPA as a real-world threat remains limited. Most OPA studies operate under idealized digital settings, where pixel values can be manipulated with exact precision and directly fed into the model. In contrast, real-world scenarios involve multiple sources of variability, including sensor noise, compression artifacts, illumination changes, and geometric transformations. These factors can easily disrupt single-pixel perturbations, significantly reducing their effectiveness outside controlled environments. Additionally, several studies have shown that simple pre-processing techniques such as image compression, resizing, or denoising can mitigate or completely neutralize one-pixel perturbations. This suggests that OPA may not pose a substantial standalone threat in many practical systems, particularly those incorporating basic input sanitization pipelines. Nevertheless, OPA retains significant importance from a theoretical and diagnostic perspective. It serves as an extreme case of ultra-sparse adversarial perturbations, providing insight into the fragility of model representations and the existence of highly localized vulnerabilities. Additionally, OPA has inspired a broader class of sparse and few-pixel attacks that are more robust and practically relevant. In this context, OPA should be interpreted not as a direct real-world attack vector, but as a foundational tool for probing model sensitivity and guiding the development of more realistic adversarial frameworks. Future research should therefore focus on bridging the gap between theoretical minimal perturbations and physically realizable adversarial scenarios.

### 9.8. Future Directions

The trajectory of OPA research must shift from empirical demonstrations of vulnerability to the creation of fundamentally resilient AI systems. This requires a departure from surface-level patches and toward theoretical explorations of model fragility and adversarial susceptibility. Proactive defence strategies should be embedded into the core design of neural networks, leveraging insights from AI security, privacy, and ethics to formulate comprehensive and anticipatory solutions. Real-world relevance must be a guiding principal models and defences need to be tested against dynamic and noisy conditions, not just pristine academic datasets. Additionally, the field lacks standardized evaluation metrics and benchmarks specifically tailored for OPA, which hinders cross-comparability and reproducibility of results. Establishing such standards is essential for coherent progress and the development of scalable, generalizable solutions. Ultimately, a multidisciplinary, insight-driven approach will be pivotal in safeguarding the next generation of AI technologies. In light of the insights gained from this study, several key research avenues are proposed to advance the field of adversarial robustness and address fundamental vulnerabilities in DNN (DNNs):

i. A major limitation of the current OPA literature is the absence of standardized benchmarking. Most studies evaluate different optimization algorithms using different datasets, model architectures, attack objectives, and reporting metrics, preventing statistically valid cross-study comparisons. Future benchmark initiatives should evaluate multiple optimization strategies under identical experimental conditions using common datasets, standardized model architectures, and harmonized evaluation metrics to enable reproducible quantitative meta-analysis and objective assessment of algorithmic performance.

ii. Future research should emphasize integrating robustness directly into DNN architectures. Approaches such as Receptive Field Hardening can enhance resistance to localized pixel perturbations, while Spectral Regularization penalizes high-frequency feature activations during training. These techniques aim to mitigate the root causes of model fragility, enabling inherently robust networks rather than relying solely on post hoc defence mechanisms.

iii. Establishing a systematic classification of OPA failure modes is essential. By analyzing how vulnerabilities manifest across different architectures (CNNs vs. transformers) and specific image regions (object boundaries versus backgrounds), researchers can deepen their understanding of adversarial behaviour. This taxonomy would facilitate more precise and architecture-aware defence strategies.

iv. Integrating explainability techniques such as saliency maps or attribution-based methods with attack algorithms like Differential Evolution could yield more targeted and efficient adversarial examples. This hybrid approach not only enhances attack efficacy but also illuminates the pixel-level sensitivities exploited by attacks, guiding the design of more focused and informed defence mechanisms.

v. There is a pressing need for computationally efficient, interpretable lightweight detection systems suitable for deployment in resource-constrained environments. Techniques leveraging shallow classifiers or simple statistical features such as pixel intensity histograms and entropy measures can offer effective detection with minimal overhead, making them ideal for security-critical applications with limited computational capacity.

vi. Rather than applying blanket modifications across entire images, future defences should prioritize selective pixel-level restorations, especially in domains where preserving image fidelity is crucial such as medical imaging. Moreover, defences should be co-developed with domain-specific considerations in mind for instance, incorporating federated adversarial training to maintain patient privacy, or leveraging hardware acceleration for real-time defence in IoT or cyber-physical systems.

vii. The field would benefit significantly from the development of standardized frameworks for evaluating and comparing OPA techniques and defences. This includes the creation of diverse, real-world cross-domain benchmarks such as native-resolution datasets from healthcare, remote sensing, and transportation, as well as the adoption of unified evaluation metrics that accurately reflect the trade-off between attack stealth and defence efficacy.

viii. Future research should move towards modelling the attacker-defender interaction as an evolving, adaptive process. Leveraging reinforcement learning to simulate attacker strategies during defence development can enhance robustness. Additionally, explainable proactive detection combining trigger localization with saliency-based analysis could enable the identification of vulnerable regions before an attack materializes, marking a critical shift from reactive to preventive security measures.

ix. OPAs can act as "stress tests" tool to identify brittle regions in DNN. It can serve as effective diagnostic tools to identify vulnerabilities in deep learning models by revealing fragile decision boundaries with minimal perturbation. As benchmarks, they enable researchers to evaluate and compare model robustness across architectures and datasets. Additionally, OPA contribute to explainable AI by highlighting high-saliency regions, providing insights into how models prioritize input features. In security contexts, they support lightweight adversarial testing, helping ensure that models are resilient to imperceptible manipulations. Collectively, these benefits position OPA as essential instruments for developing more robust, interpretable, and trustworthy AI systems.

x. Unlike CNNs, transformers process images through patch-level attention, raising questions about their susceptibility to localized pixel perturbations. Key directions include analysing how one-pixel changes affect attention distributions and token embeddings, comparing vulnerability patterns between CNNs and transformers, and using such attacks to improve explainability through attention shift visualization.

xi. Vision Transformer (ViT)-specific vulnerability analysis represents an important research gap in the current OPA literature. Among the 32 studies included in this review, none explicitly quantified whether ViT patch tokenization provides an inherent defense against one-pixel perturbations. Existing OPA research remains heavily focused on convolutional architectures such as ResNet, VGG, DenseNet, and LeNet. Because ViTs encode images as patch-level token embeddings rather than localized convolutional receptive fields, isolated pixel perturbations may be distributed across patch representations, potentially reducing their influence on downstream classification. Nevertheless, this hypothesis has not yet been systematically evaluated in the reviewed literature. Future research should directly compare CNNs and transformer-based architectures under identical one-pixel attack settings to determine whether patch tokenization confers measurable robustness or simply alters the attack surface.

**9.9. Proposed OPA Security Regulation Policy**

OPAs, though technically small, create systemic risks when targeted at safety-critical systems; therefore, organizations and regulators should treat OPA as a cybersecurity and safety concern. Recommended measures include:

a. Mandatory adversarial-robustness testing and adversarial-testing documentation for high-risk AI systems deployed in healthcare, automotive, and critical infrastructure
b. A formal incident-reporting requirement and timeline for serious AI incidents (aligned with existing cyber incident rules) so that regulators and downstream users can assess and contain harm;
c. Compulsory vulnerability disclosure channels and CVE [125]/CWE [126] registration for significant model vulnerabilities;
d. Explicit liability regimes where vendors and integrators of high-risk AI systems bear product-liability and compliance responsibilities if insufficient adversarial testing or mitigation was performed; and
e. Clear criminal and civil remedies for intentional misuse combined governance (standards, best practices and certification) and legal enforcement will align incentives toward robust design and timely remediation.

Regulators should recognize serious model-integrity incidents as a distinct class of cybersecurity events and implement a balanced regime that combines mandatory safeguards for high-risk systems with strong incentives for responsible research and disclosure. The proposed policy framework builds on established legal foundations such as the robustness and cybersecurity obligations in the EU AI Act, the incident-reporting requirements of the GDPR [120], and operational guidance from national cybersecurity bodies like NIST and CISA adapting them to the context of model-integrity threats. It outlines seven key components: (1) defining adversarial machine learning as a model-integrity threat and classifying incidents by risk severity; (2) enforcing standardized robustness testing for high-risk AI systems against both dense and sparse perturbations; (3) requiring 72-hour incident reporting to designated authorities detailing impacts and mitigation steps; (4) establishing coordinated vulnerability disclosure and public CVE-style registries; (5) providing safe-harbour protections for researchers following responsible-disclosure protocols; (6) imposing liability on vendors that deploy inadequately tested models; and (7) aligning all practices with international frameworks, including the EU AI Act [119](Article 15), the NIST AI Risk Management Framework [118], and the Budapest Convention [117]. Collectively, these measures strengthen transparency, accountability, and resilience throughout the adversarial machine-learning ecosystem. **Figure 19** represents the Regulatory Governance Model for Adversarial ML Incidents (for OPA context)

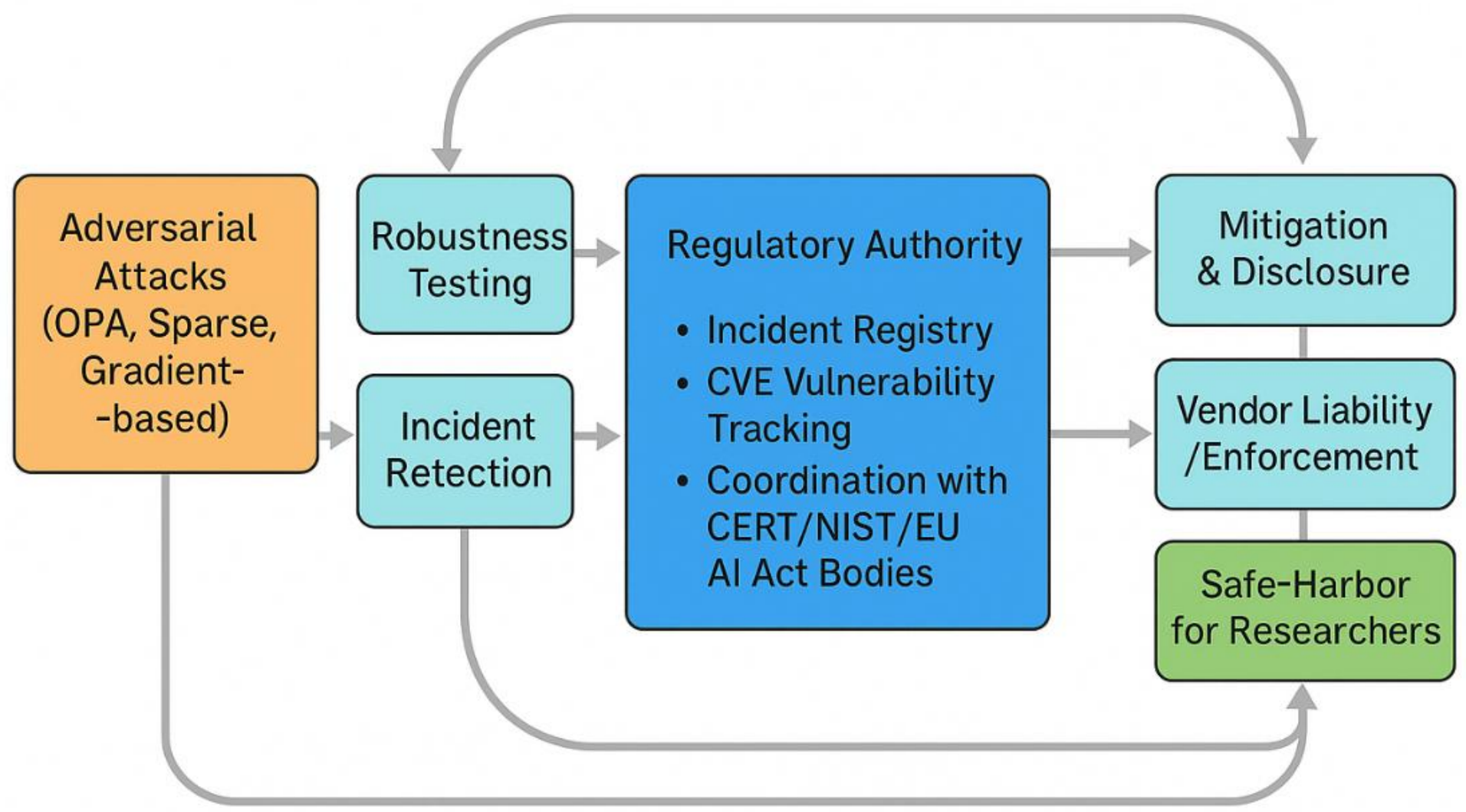

**Figure 19.** Regulatory Governance Model for OPA

From a regulatory perspective, vulnerability to one-pixel attacks should be interpreted within a risk-based assessment framework rather than as an automatic indicator of deployment failure. Although OPA provides a valuable benchmark for evaluating decision-boundary sensitivity under controlled digital conditions, current evidence suggests that many single-pixel perturbations become substantially attenuated in physical imaging pipelines because of optical blur, sensor noise, image compression, and environmental variability. Consequently, AI deployment vendors should not be considered inherently liable for susceptibility to theoretical digital attacks that do not translate into credible real-world threats. Instead, regulatory requirements should align with the intended deployment environment. Digital-only applications, such as cloud-based medical image analysis or forensic image processing, should demonstrate robustness against digitally realizable adversarial perturbations, whereas camera-based systems should additionally be evaluated using physically realizable attack scenarios. This risk-proportionate approach enables OPA to remain an important standardized robustness benchmark while avoiding disproportionate regulatory burdens for vulnerabilities that have limited practical exploitability.

## 10. Conclusion

OPAs expose a critical vulnerability in deep neural networks by demonstrating that imperceptible, single-pixel perturbations can cause high-confidence misclassifications. This PRISMA-guided review synthesized 32 key studies and introduced a structured taxonomy spanning attack strategies, defence mechanisms, interpretability tools, and domain-specific applications. Our analysis highlights the dominance of black-box optimization techniques, the limitations of current defence approaches, and the underutilization of interpretability in both attack design and robustness evaluation. As the field evolves, future research must prioritize lightweight black-box detection, explainability-driven strategies, and integrated, domain-adaptive defences. From a policy standpoint, regulators should establish standardized robustness testing protocols, formal incident-reporting mechanisms for AI integrity breaches, and coordinated vulnerability disclosure frameworks supported by researcher safe-harbor provisions. Clear accountability guidelines and cross-sector compliance alignment with frameworks such as the EU AI Act, NIST AI RMF, and Budapest Convention will help ensure responsible AI governance. Although this review demonstrates substantial evidence that one-pixel attacks expose important vulnerabilities in CNN-based computer vision systems, comparatively little quantitative evidence currently exists for frontier architectures such as Vision Transformers, Swin Transformers, Masked Autoencoders, and multimodal vision-language models. Therefore, the conclusions of this review should primarily be interpreted within the context of the currently available literature. Expanding empirical evaluations to these modern architectures represents one of the most important directions for future adversarial robustness research. Strengthening adversarial resilience through interpretable and practical solutions is essential to ensuring trustworthy deployment of AI systems in real-world, safety-critical applications.

## Declarations

**Funding**

No Funding Available

**Declaration of Competing Interest**

The authors declare that they have no known competing financial interests or personal relationships that could have appeared to influence the work reported in this paper.

**Availability of data and materials**

Data sharing is not applicable to this article as no new data were created or analyzed in this study.

**Acknowledgements**

Not applicable